\documentclass[final,5p]{elsarticle}
\usepackage{graphicx}
\usepackage{amsmath}

\begin{document}

\begin{frontmatter}
\title{Domain-wall-assisted giant magnetoimpedance of thin-wall ferromagnetic nanotubes}
\author[PWr]{Andrzej Janutka}
\author[PWr]{Kacper Brzuszek}
\address[PWr]{Department of Theoretical Physics, Faculty of Fundamental Problems of Technology, Wroclaw University of Science and Technology, 50-370 Wroc{\l}aw, Poland}

\begin{abstract}
We study the efficiency of the magnetoimpedance (MI) of thin-walled circumferentially-ordered nanotubes 
 in sub-GHz and GHz frequency regimes, using micromagnetic simulations. 
 We consider empty ferromagnetic tubes as well as tubes filled with non-magnetic conductors 
 of circular cross-section (nanowire coverings), focusing on the low-field regime of MI (below  
 a characteristic field of the low-frequency ferromagnetic resonance). In this field area,
 the efficient mechanism of MI is related to oscillations of the positions of 
 (perpendicular to the tube axis) domain walls (DWs).
 Two mechanisms of driving the DW motion with the AC current are taken into account;
 the driving via the Oersted field and via the spin-transfer torque.
 The simulations are performed for Co nanotubes of the diameter of 300nm.
 Achievable low-field MI exceeds 100$\%$, while the field region of a high sensitivity
 of that DW-based giant MI is of the width of tens of kA/m. The later is widely adjustable
 with changing the density of the driving AC current, its frequency, and the nanotube length.
 Of particular interest is the resonant motion of DW due to the 
 interaction with the nanotube ends, the conditions of whom are discussed.
\end{abstract}

\end{frontmatter}

{\small PACS R}ef: 75.78.Cd,75.78.Fg,76.50.+g,85.70.Kh\\
%\newpage
\footnotemark{E-mail address: Andrzej.Janutka@pwr.edu.pl}

\section{Introduction}
The giant magnetoimpedance (GMI) is utilized in sensing weak magnetic fields (e.g. Oersted fields,
 geomagnetic fields, stray fields, etc.) with a very large field sensitivity (compared to the one
 achievable with the giant magnetoresistance or the Hall effect) while with low requirements
 on working conditions of the sensor (contrary to the SQUID magnetometers) \cite{kno03,pha08,zhu09,zhu16}.
 Though, magnetic systems of different geometries are considered to be active ingredients
 of the GMI-based sensors, the conducting ferromagnetic cylinder (the ferromagnetic wire) 
 and the tube are base systems when building mathematical models of GMI.
 Within these two, the tube geometry is attractive because it allows for 
 better isolating different mechanisms of the magnetoimpedance (MI) than
 a magnetic wire. The wires that reveal the GMI (e.g. amorphous
 magnetic microwires with negative constant of the volume magnetostriction)
 are of a complex magnetic structure, with an axially-magnetized inner core
 and a circumferentially-magnetized outer shell, whereas, the inner core is absent in the tube.
 This removes one of the basic mechanisms of the (middle-frequency) MI of the wires which
 is based on the dependence of the inner-core radius on the external axial field,
 (the operating frequency for that MI mechanism is limited by the skin effect). 
 In the low-frequency range, that mechanism of MI of the magnetic wire coexists
 with an impedance-inducing oscillatory motion of the domain walls (DWs) in the other shell.
 In order to isolate the later (low-frequency) mechanism of MI, one has proposed to use
 a conductive non-magnetic microwire covered with a tubular ferromagnetic layer. 
 An advantage of the tube geometry with regard to the high-frequency MI, 
 whose mechanism is based on the ferromagnetic resonance (FMR), follows from a magnetic softening
 of the tubes relative to the wires \cite{kur11}, (in the microwires, the magnetostatic 
 core-shell interaction can be strong, especially, in the case of the circumferential ordering of 
 the outer shell \cite{zhu18}).
 
Though, the middle- (skin-effect-based) and high-frequency (FMR-based) GMI of the soft-magnetic microwires
 and microtubes are very-well-described and optimized effects, (the thresholds are typically in sub-MHz and sub-GHz 
 ranges), DW-based MI is not explored in detail.
 A huge limitation of its efficiency in the micro-sized systems follows from
 the damping of the Oersted-field-driven motion of the DW by eddy currents which this motion
 induces \cite{atk98,zim15}. The damping is expected to strengthen with the driving-current frequency,
 while its details are under debate, (note that performing full micromagnetic simulations
 of the microwires/microtubes remains to be a challenge), \cite{val02,pan04,ciu07}. 
 In the nanotubes of the wall thickness
 comparable to the electronic mean free path, such eddy currents are expected to be suppressed. 
 Recent development of the techniques of the nanowire and nanopore coating
 (atomic-layer deposition, electroplating) and rolling up the nanomembranes
 allows for creating the tubes of extremely small thickness of the wall, whose domain structures
 are stable in wide ranges of external conditions \cite{cho10,web12,str14,gro16,wys17,sta17}.
 Via adjusting parameters of the manufacturing process (oblique-evaporation
 and thermal treatment, \cite{van01,che07}) and material composition \cite{esc07}, the magnetic anisotropy
 can be largely tuned allowing for stabilization of the circumferential ordering in the nanotube. 
 Such an ordering is crucial for GMI, however, it has also attracted an attention with regard
 to creating a racetrack memory with a reduced (compared to the nanowires) stray field
 of (densely packed) DWs \cite{sta17}. Our study of the nanotubes is motivated by the idea
 of GMI-based sensing with a nm-sized spatial resolution \cite{fer17}.
 
There is a small number of reports on GMI of the magnetic nanosystems
 with flux-closure geometries (e.g. multi-segmented; "barcode" nanowires \cite{chi09}, 
 multilayers with closed magnetic path \cite{kur04}, nanotubes \cite{nak14}), whereas, a comparative study shows
 that property of the nanosystem to be desired for optimizing MI above some (relatively-low) threshold
 frequency of the current \cite{fer10}. While the mechanism of GMI in the barcode nanowires
 relates to the magnetization rotation in their whole volume, the DW motion plays a role in the low-field MI
 of the flux-closure multilayers whose generic model is the nanotube. 
 Note that scaling the soft-ferromagnetic microwires down to the nm-sizes leads to their 
 magnetostatically-induced hardening; vanishing the circumferentially-magnetized outer-shell. In the consequence,
 MI of the magnetic nanowires is weak \cite{ata13}, while the nanotube remains the simplest 
 high-symmetry system for studying DW-based GMI at the nanoscale. 

\begin{figure} 
\unitlength 1mm
\begin{center}
\begin{picture}(87,27)
\put(0,-4){\resizebox{87mm}{!}{\includegraphics{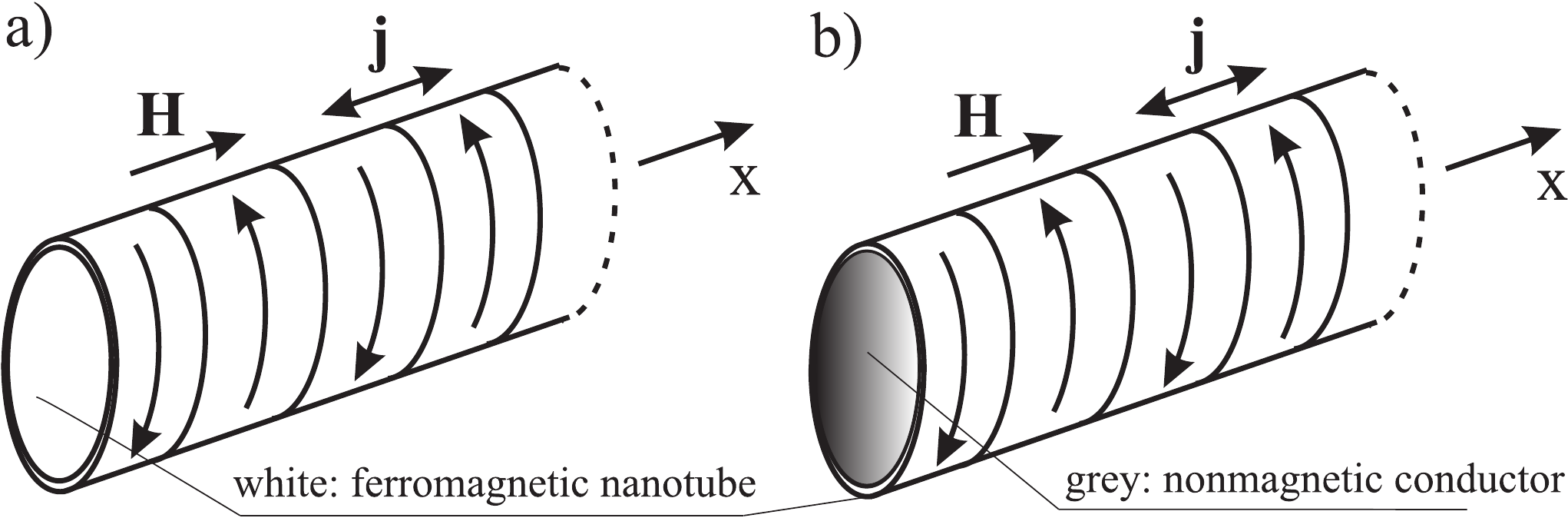}}}
\end{picture}
\end{center}
\caption{Schemes of the systems studied; the empty ferromagnetic nanotube (a), the magnetic covering 
of a nonmagnetic conducting wire (b), including the bamboo-like magnetic structure
of circumferentially-ordered domains (the curved arrows indicate the magnetization),
and indicating the direction of the external field and the electrical-AC flow.}
\end{figure} 
 
In the present paper, with micromagnetic simulations, we study the dynamical response of the magnetic 
 subsystems of DW-containing Co nanotubes to the AC current, in the presence of a constant axial magnetic field.
 We consider a tubular covering of a conducting nanowire as well as an empty nanotube (Fig. 1). 
 The tubes are of a strong easy-plane anisotropy (the circumferential ordering) and we pay especial attention
 to the low-field regime, thus, focusing on the effects of the DW motion. The time dependence of 
 the average circumferential component of the magnetization is analyzed with dependence on the current parameters
 (density amplitude, frequency) and the field and on the nanotube length, while details of the corresponding
 magnetization dynamics are monitored. We establish operating regimes of nanotube GMI.  
 
Upon introducing a model of the magnetic nanotube dynamics in Section 1, in Section 2, 
 we present the results of the micromagnetic simulations of Co nanotubes under the electrical AC current.
 Conclusions are formulated in Section 4. 

\section{Model}

\subsection{Equation of motion}

Our micromagnetic model of a thin-wall nanotube is a reduced version of a model formulated
 in \cite{jan16}. Its magnetization evolves according to the Landau-Lifshitz-Gilbert (LLG) equation
 in 3D that takes the form 
\begin{eqnarray}
-\frac{\partial{\bf m}}{\partial t}&=&\frac{2\gamma A_{ex}}{M_{s}^{2}}{\bf m}\times\Delta{\bf m}
\nonumber\\
&&+\gamma{\bf m}\times({\bf B}_{ms}+{\bf B}_{Oe}+{\bf B})
\nonumber\\
&&-\frac{\eta j}{M_{s}^{2}}{\bf m}\times\left({\bf m}\times\frac{\partial{\bf m}}{\partial x}
+\beta M_{s}\frac{\partial{\bf m}}{\partial x}
\right)
\nonumber\\
&&+\frac{2\gamma K}{M_{s}^{2}}({\bf m}\cdot\hat{i}){\bf m}\times\hat{i}
-\frac{\alpha}{M_{s}}{\bf m}\times\frac{\partial{\bf m}}{\partial t}.
\label{LLG}
\end{eqnarray}
Here, $\hat{i}\equiv[1,0,0]$, (the wire is directed along the $x$ axis), $M_{s}=|{\bf m}|$
 represents the saturation magnetization, $A_{ex}$ denotes the exchange stiffness,
 $\gamma$ - the gyromagnetic factor), $K$ determines the strength of the effective axial
 anisotropy (specified in the next paragraph), $\alpha$ - the Gilbert damping constant.
 The magnetostatic and external fields are denoted by ${\bf B}_{ms}$
 and ${\bf B}$, respectively. The presence of the electric current of the density $j$ 
 (that flows along the nanotube) is included via the Oersted-field (${\bf B}_{Oe}$)
 term on the right-hand side of (\ref{LLG}) and via the spin-transfer torque (STT); whose coupling constant
 $\eta=Pg\mu_{B}/2eM_{s}$ is dependent on the spin polarization of the system $P$, the Lande factor $g$,
 the Bohr magneton $\mu_{B}$, and the absolute value of the electron charge $e$. The unitless parameter 
 $\beta$ determines the strength of the non-adiabatic STT. Since there is no commonly-accepted methodology
 for measuring $\beta$, \cite{bea08}, we follow a (structure-stability-based) reasoning of \cite{tse08}
 that leads to $\beta\approx\alpha$. Let us note that performing micromagnetic simulations
 with the simplified (Landau-Lifshitz) form of the LLG equation requires rescaling the input
 parameter $\beta\to2\beta$.

Here, we restrict our considerations to the circumferentially-magnetized tubes,
 which requires the axial-anisotropy constant $K$ to be negative and dominate 
 over the shape anisotropy of the magnetostatic origin.
 The main contributions to the anisotropy have been described in \cite{jan16}
 to relate to the internal stress of the tube that is dependent on the manufacturing 
 method. With relevance to the thin-wall microtubes, we have specified two basic regimes
 of the anisotropy which correspond to a solidification-dominated stress and
 a cooling-dominated stress. The later drives the circumferential ordering 
 in the tube (wire) made of a material with a negative constant of the 
 volume magnetostriction. The former can be induced by specifying the direction
 of the material deposition when forming the tube-shaped magnetic layer (via a "shadowing"
 effect, i.e. the creation of a crystallographic texture, and/or interface steps) \cite{van01,che07}.
 In the tubes of ultimately-thin walls, such a solidification-induced anisotropy 
 is expected to determine the ordering direction.

The (circumferential) Oersted field is dependent on the radial coordinate
 of the tube $\rho\equiv\sqrt{y^{2}+z^{2}}$, [let the angular coordinate be
 $\varphi\equiv{\rm arctan}(z/y)$].
 In the case of the empty nanotube, assuming the current density to be uniform, via
 the Ampere's law, one finds the circumferential field (the transverse component of the field) 
 $H_{\varphi}(\rho,t)=-j(t)(\rho^{2}-R_{in}^{2})/2\rho$,
 where $R_{in(out)}$ denotes the inner (outer) radius of the tube.
 When the tube is considered as a covering of a circular nanowire, the Oersted field
 is taken in the form $H_{\varphi}(\rho,t)\approx\bar{H}_{\varphi}(t)=-j(t)\bar{R}/2$, 
 with $\bar{R}=(R_{in}+R_{out})/2$. Averaging the Oersted field over the radial
 coordinate of the tube is performed following
 $\bar{H}_{\varphi}=(R_{out}-R_{in})^{-1}\int_{R_{in}}^{R_{out}}H_{\varphi}(\rho,t){\rm d}\rho$.
 For a given value of the current density, the Oersted 
 field in the covering of the nanoconductor is considerably (several times) higher 
 than for the case of the empty nanotube, whereas, the STT amplitude is independent of the 
 tube filling. 
 
\subsection{Dynamical regimes}

The MI of thin-wall nanotubes is expected to follow from the DW motion for up
 to GHz frequencies of the current, whereas, the importance of the DW motion for the microwire or microtube GMI
 is believed to be restricted to a low-frequency (kHz) regime \cite{gar01,kra03}. 
 We perform a simple estimation of the upper frequency limit of DW-based MI, noticing that 
 the second of two basic mechanisms of GMI in the nanotube (a high-frequency one) relates to exciting FMR.
 Considering sufficiently low axial fields ($H_{x}\ll M_{s}$), we claim the frequency of the zero-field FMR
 to be that upper bound. We evaluate it applying the Kittel-like formula for the FMR frequency, \cite{kra03,kra99,kit48},
\begin{eqnarray} 
\omega_{R}(H_{x})=\gamma\mu_{0}\sqrt{M_{s}\left[H_{x}\cos(\Theta)+\frac{2|K|}{\mu_{0}M_{s}}\cos(2\theta)\right]},
\label{FMR}
\end{eqnarray}
where $M_{s}$ denotes the saturation magnetization, $K$ - the constant of axial anisotropy; $K<0$,
 $\mu_{0}=1.26\cdot10^{-6}$N/A$^{2}$ - the vacuum permeability,
 and $\gamma\mu_{0}=2.21\cdot10^{5}$m/sA - the gyromagnetic factor.
 The angles $\theta$ and $\Theta$ measure the deviation of the magnetization 
 from the static external field and from the plane of the tube cross-section, respectively,
 (here $\Theta=\pi/2-\theta$). Using the material parameter of Co ($M_{s}=1400$kA/m),
 for the nanotube with axial hard-axis anisotropy of the constant $K=-150$kJ/m$^{3}$,
 (a systems of the diameter of 300nm studied in \cite{jan16}), at $H_{x}=0$, we find the resonance
 frequency to be as large as $\omega_{R}(0)=2\pi\cdot17$GHz. Our micromagnetic simulations
 of the current-driven motion of DWs in the relevant Co nanotubes (Section 3) are restricted to 
 the AC frequencies well below 17GHz, thus, the effect of finite width of the FMR peak
 is avoided in the vicinity of $H_{x}=0$. 

A typical curve of the dependence of the resonance frequency
 on the axial field is given in Fig. 2, (here $H_{K}\equiv2|K|/\mu_{0}M_{s}$). 
 The field regime I corresponds to the stability of the domain structure (provided the current
 frequency is considerably smaller than the resonance one), and it is relevant to DW-based MI.  
 In the regime II, the domain-structure is absent and FMR-based MI is observed.
 In a transient regime of the field, (a dashed area in Fig. 2), there is an overlap of the driven motion
 of DWs with FMR, which results in an instability of the domain structure and of the impedance. Within 
 that regime, there are areas of dominance of the DW motion over FMR (IIIA; the effect of finite spectral width 
 of the resonance) and of the dominance of the FMR over an evolution of the domain structure
 (IIIB; the effect of incomplete relaxation of the uniform state due to the DW motion).
 
\begin{figure} 
\unitlength 1mm
\begin{center}
\begin{picture}(85,47)
\put(0,-4){\resizebox{85mm}{!}{\includegraphics{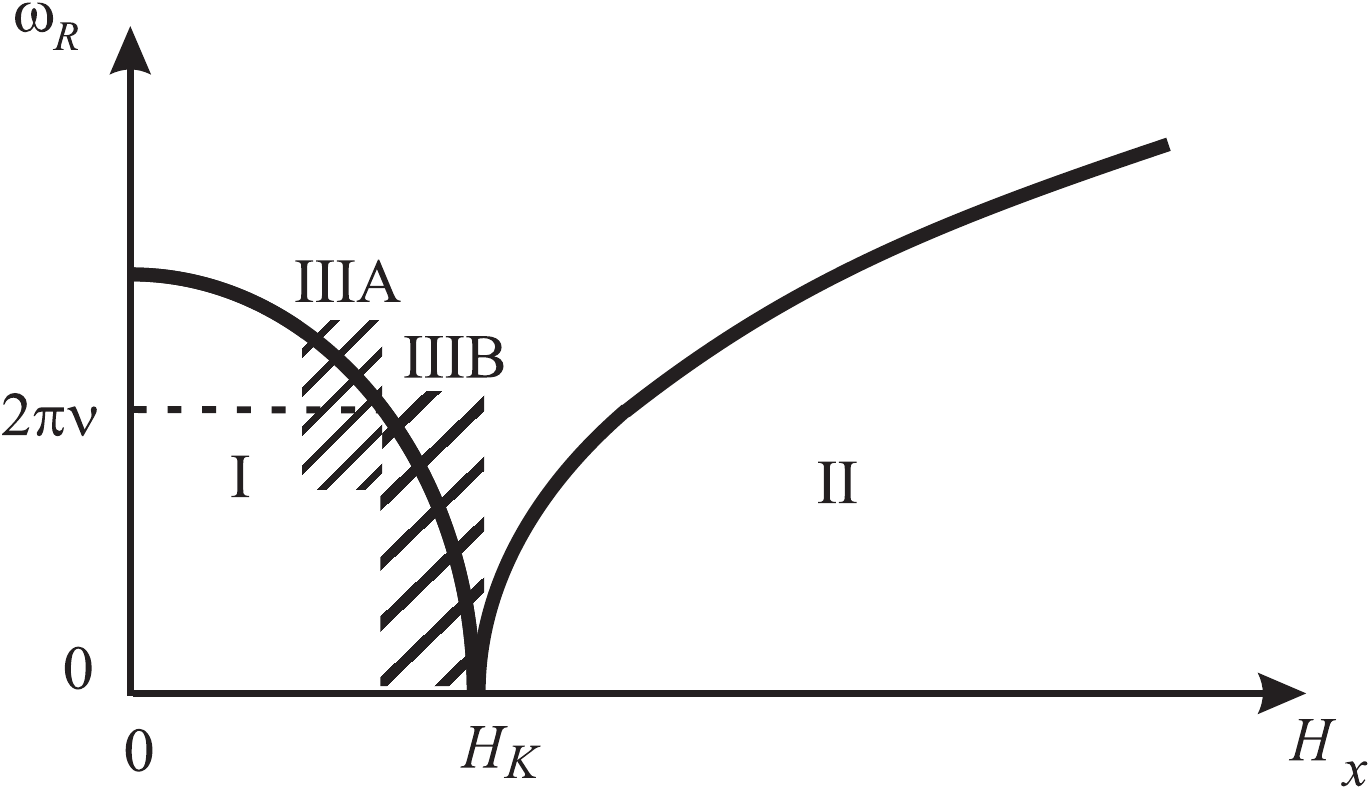}}}
\end{picture}
\end{center}
\caption{The dependence of the frequency of FMR in a thin film on the in-plane field perpendicular to 
the driving field in the easy (in-plane) direction. The areas I and II correspond to the domain-structure
stability and the single-domain state, respectively, while IIIA and IIIB to the transient regimes
of the chaotic and regular dynamical response to the alternating (Oersted) field, respectively.}
\end{figure} 

Large value of the anisotropy constant is crucial for the above estimation, while, it is close to the minimum
 required in order to ensure the stability of the circumferential ordering in our system. As mentioned above,
 the anisotropy constant is not any universal parameter since the anisotropy originates from the method of fabrication
 of the nanotube and from the curvature of its layer.
 However, the above taken value for Co nanotube is realistic (several times larger
 than reported in Ref. \cite{web12} for Ni nanotubes of similar diameter as the tubes simulated in the present work
 while of lower saturation magnetization \cite{dau07}, an order of magnitude larger than for rolled-up Co nanomembrane
 of the diameter of 5$\mu$m \cite{mul09}, and an order of magnitude larger than for CoNiB nanotubes
 of a higher thickness of the wall \cite{sta17}). 

Finally, let us verify the negligibility of the skin effect, estimating the skin depth for cobalt with
 $\delta_{Co}^{-1}=\sqrt{\pi\nu\sigma_{Co}\mu_{Co}\mu_{0}}$, where we substitute; the electrical resistivity
 $\sigma_{Co}^{-1}=6.2\cdot10^{-8}\Omega$m, the relative permeability $\mu_{Co}=250$, and the maximum current frequency
 of our micromagnetic simulations $\nu=6$GHz. We obtain the lower bond on the skin depth; $\delta_{Co}\ge33$nm, 
 which is three-time larger than the wall thickness of the simulated nanotubes.

\subsection{Domain wall under AC current}

The static DW solutions of the present model have been established in \cite{jan16}. 
 With the purpose of establishing the frequency window of the stable oscillatory motion 
 of DWs under the AC current, a crucial problem to discuss is the DW inertia. 
 According to the Walker model of the DW motion, the inertia results
 from the field- or current-driven deviation of the DW magnetization from its equilibrium
 orientation. Since this way excited DW is not static while it propagates, the time of 
 the relaxation of its magnetization is also a time of braking the translational motion upon
 switching the external field off. That DW relaxation time $\tau_{1D}$ can be determined
 from the equations of motion of the collective coordinates of DW (\ref{q-phi}); the position
 of the DW center and the angle of deviation of its magnetization from the static orientation.
 According to \cite{thi07}, the relaxation rate takes the following dependence on an effective
 in-plane anisotropy field $H_{M}$
\begin{eqnarray}
\Gamma=\tau_{1D}^{-1}=\frac{\alpha\gamma\mu_{0}H_{M}}{1+\alpha^{2}}.
\end{eqnarray}
Since the in-plane anisotropy field is dominated by the stray-field contribution; $H_{M}=M_{s}-H_{K}\approx M_{s}$,
 for a thin-wall Co nanotube, with $\alpha=0.05$, one estimates $\tau_{1D}=0.4$ns. 
 Note that, in relevant evaluations for soft-magnetic nanostripes \cite{bea08,thi06,jan11},
 $H_{M}=2H_{W}/\alpha$, where $H_{W}$ denotes the critical field of the Walker breakdown,
 while, following \cite{sch74}, for simple DWs (without vortex-like singularities),
 $H_{W}=\alpha M_{s}/2$, thus, the relation $H_{M}\approx M_{s}$ is valid,
 (in the presence of singularities inside DW; $H_{W}<\alpha M_{s}/2$, \cite{bea08}). 
 
When inducing MI in long nanotubes via oscillations of the DW position (the interaction
 of DW with the tube ends is negligible), the relaxation rate
 $\Gamma$ must be smaller than the circular frequency, (the underdamped oscillations regime),
 which gives the lower bound on the AC frequency $\nu>\Gamma/2\pi=2.5$GHz.
 
\begin{figure*} 
\unitlength 1mm
\begin{center}
\begin{picture}(175,170)
\put(24,-7){\resizebox{144mm}{!}{\includegraphics{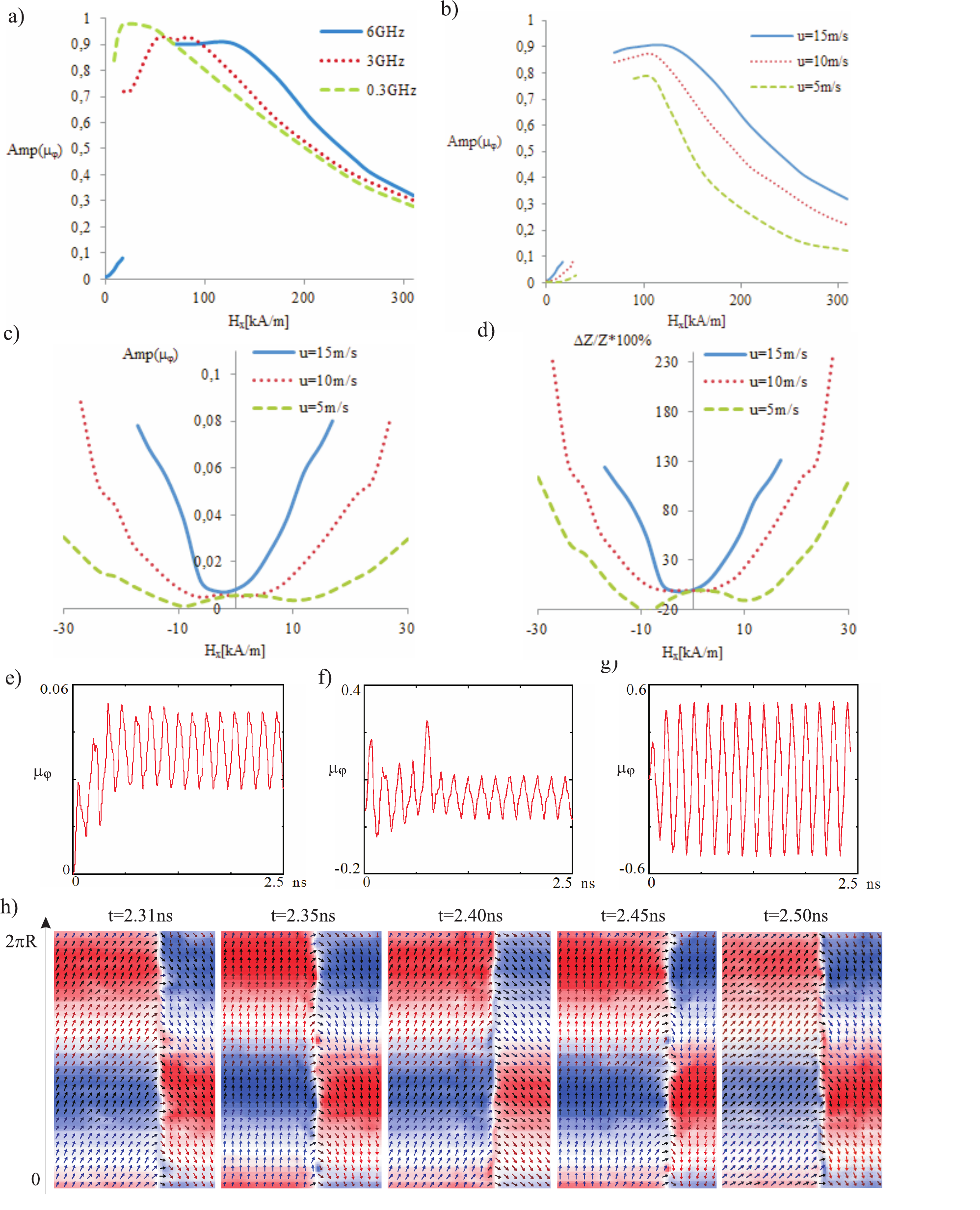}}}
\end{picture}
\end{center}
\caption{Simulations of 3$\mu$m-long tubular Co covering of the conducting nanowire of the outer diameter
of 300nm and the wall thickness of 11nm, relaxed from the uniform magnetization state ${\bf m}=(M,0,0)$. 
The amplitude of oscillations of the summary circumferential magnetization 
of the nanotube (a)-(c) and the relative impedance (d) with dependence on the longitudinal field. 
In (a), the amplitude of the scaled current density is u=15m/s, while, in (b) and (c), the current frequency is 6GHz.
Time dependence of the summary circumferential magnetization for the current of the frequency of 6GHz
and the scaled density u=15m/s at the longitudinal field 
$H_{x}=3$kA/m (e), $H_{x}=15$kA/m (f), $H_{x}=250$kA/m (g). Snapshots of the central 500nm-long area
of the nanotube (h) which correspond to (f), (a single period of the DW evolution).
The area colors and their intensity correspond to $y$-component of the magnetization.}
\end{figure*} 

The effect of finite length of the nanotube provides a binding potential dependent
 on the DW position, (thus, an eigenfrequency of the position oscillations, Appendix A). 
 It distinguishes a frequency area around the point of the resonance oscillations
 of the DW position below the above-established upper bound on frequency range
 of DW-based MI $\nu<\omega_{R}(0)/2\pi$ ($=17$GHz for our nanotubes of Co).
 The borders of that near-resonance range can be obtained from the condition
 of a small shift of the oscillator phase relative to the phase of the driving field,
 (smaller than arbitrarily chosen $\pi/4$ value, \cite{bea08,thi06}). 
 From (\ref{second-order-q}), this condition is written with
\begin{eqnarray}
-1<\frac{2\pi\nu\Gamma}{\omega_{0}^{2}-(2\pi\nu)^{2}}<1, 
\label{critical_inequality}
\end{eqnarray}
where, $\omega_{0}^{2}=\omega_{M}^{2}2\delta\Delta/L$
 and $\Gamma=\omega_{M}\alpha(1+2\delta\Delta/L)$, with
 $\omega_{M}\equiv\gamma\mu_{0}M_{s}$.
 The solution to (\ref{critical_inequality}) takes the form
\begin{eqnarray}
2\pi\nu>\sqrt{\omega_{0}^{2}+\frac{\Gamma^{2}}{4}}+\frac{\Gamma}{2}\equiv2\pi\nu_{c2}\hspace*{0.5em};\hspace*{0.5em}(2\pi\nu)^{2}>\omega_{0}^{2},
\nonumber\\
2\pi\nu<\sqrt{\omega_{0}^{2}+\frac{\Gamma^{2}}{4}}-\frac{\Gamma}{2}\equiv2\pi\nu_{c1}\hspace*{0.5em};\hspace*{0.5em}(2\pi\nu)^{2}<\omega_{0}^{2}.
\end{eqnarray}
For a Co tube of the length of 3$\mu$m and of the DW width $\Delta\approx15$nm, 
 ($\Delta\approx l_{K}\equiv\sqrt{A_{ex}/|K|}$ following Appendix A, while 
 $A_{ex}=3.3\cdot10^{-11}$J/m, $K=-1.5\cdot10^{5}$J/m$^{3}$), in the limit of thick-wall
 tube (a nanowire); $\delta\to1$, the critical frequencies $\nu_{c2}=6.2$GHz and $\nu_{c1}=3.9$GHz are the minimum
 and maximum frequencies of the regimes of stable "in-phase" oscillations of the DW position, respectively.
 Note that the presence of the external longitudinal field influences the DW width \cite{jan11}, thus,
 shifting the eigenfrequency of the oscillator $\omega_{0}$ and the critical frequencies
 $\nu_{c1}$, $\nu_{c2}$ upward.
 In the limit of thin-wall tube; $\delta\to0$, one finds $\nu_{c1}=0$ and $\nu_{c2}=\Gamma/2\pi=2.5$GHz,
 which result is almost independent of the tube length. Hence, for the nanotube of the ultimately-thin wall,
 the frequency area of the resonance oscillations of the DW position is bordered by the area of the underdamped 
 oscillations, (the resonant oscillations are overdamped). In order to see the effect of resonance
 of the DW position in the simulations, we include one frequency point of the sub-GHz range in our numerical studies.  
 
\section{Micromagnetic simulations}

Before simulating systems of many DWs, we study the AC driven 
 motion of a single DW in a nanotube and its relationship to MI. 
 We have performed a series of micromagnetic simulations (using OOMMF package \cite{oommf}) of two basic systems;
 the tubular magnetic
 covering of a nanowire and the empty nanotube. The qualitative difference in modeling both follows from
 different distributions of the Oersted field. In the initial state, the DW is placed in the center of the nanotube.
 According to \cite{jan16}, such a state is created via the relaxation of the tube (with a hard-axis anisotropy)
 from the state of the uniform magnetization directed along the tube axis (then the Neel DW is created 
 in the tube center) or perpendicular to the tube axis (then the cross-tie DW is created).

\begin{figure*} 
\unitlength 1mm
\begin{center}
\begin{picture}(175,76)
\put(24,-7){\resizebox{150mm}{!}{\includegraphics{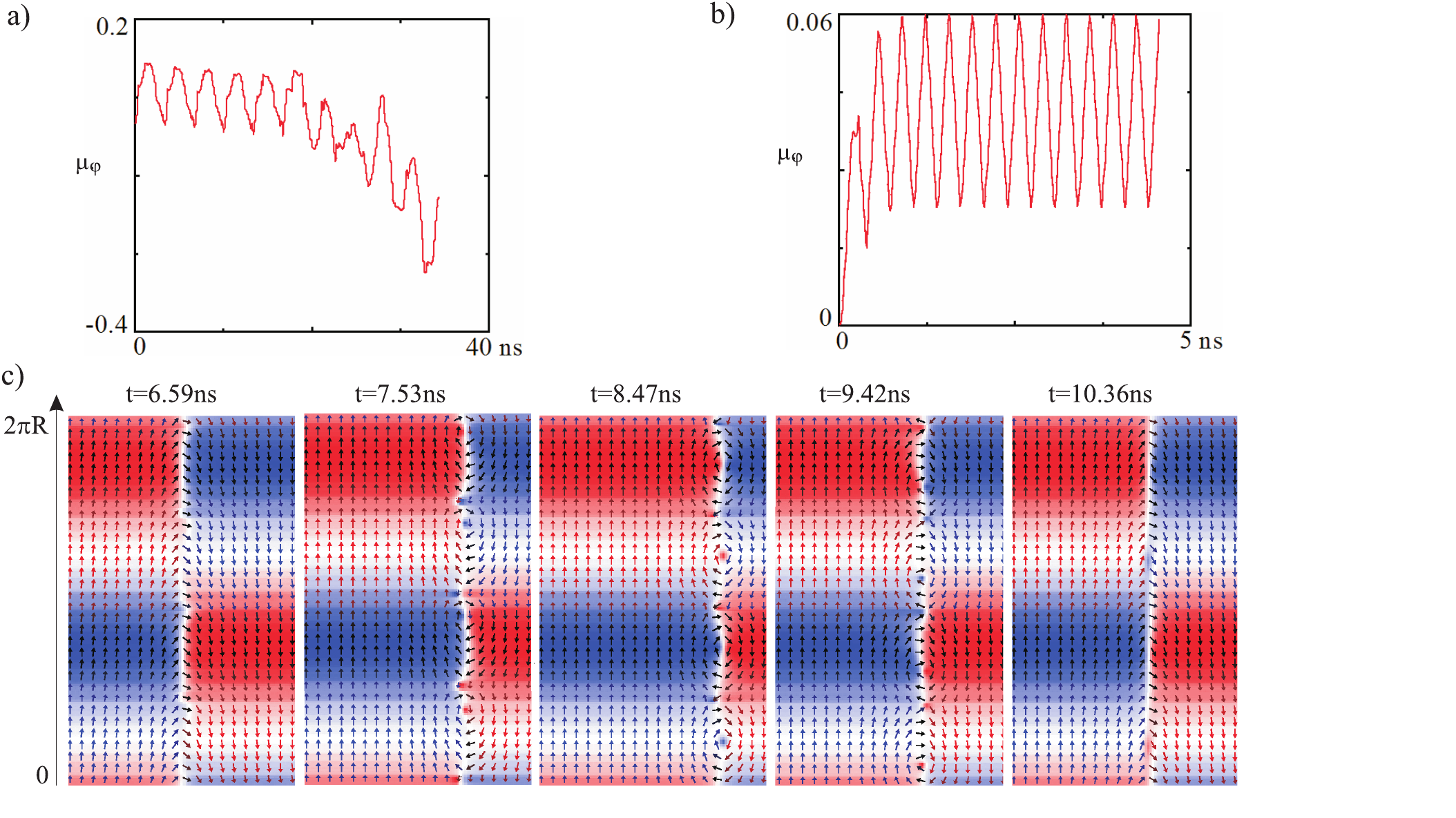}}}
\end{picture}
\end{center}
\caption{Simulations of 3$\mu$m-long tubular Co covering of the conducting nanowire (300nm diameter, 11nm thickness)
relaxed from the uniform magnetization state ${\bf m}=(M,0,0)$ under the AC current of the scaled-density amplitude $u=5$m/s.
Time dependence of the summary circumferential magnetization for the current of the frequency of 0.3GHz (a) and 3GHz (b)
at zero longitudinal field, and snapshots of the central 500nm-long area of the nanotube with the current 
of frequency 0.3GHz (c) which correspond to (a).
The area colors and their intensity correspond to $y$-component of the magnetization.}
\end{figure*}

\begin{figure*} 
\unitlength 1mm
\begin{center}
\begin{picture}(175,84)
\put(24,-7){\resizebox{136mm}{!}{\includegraphics{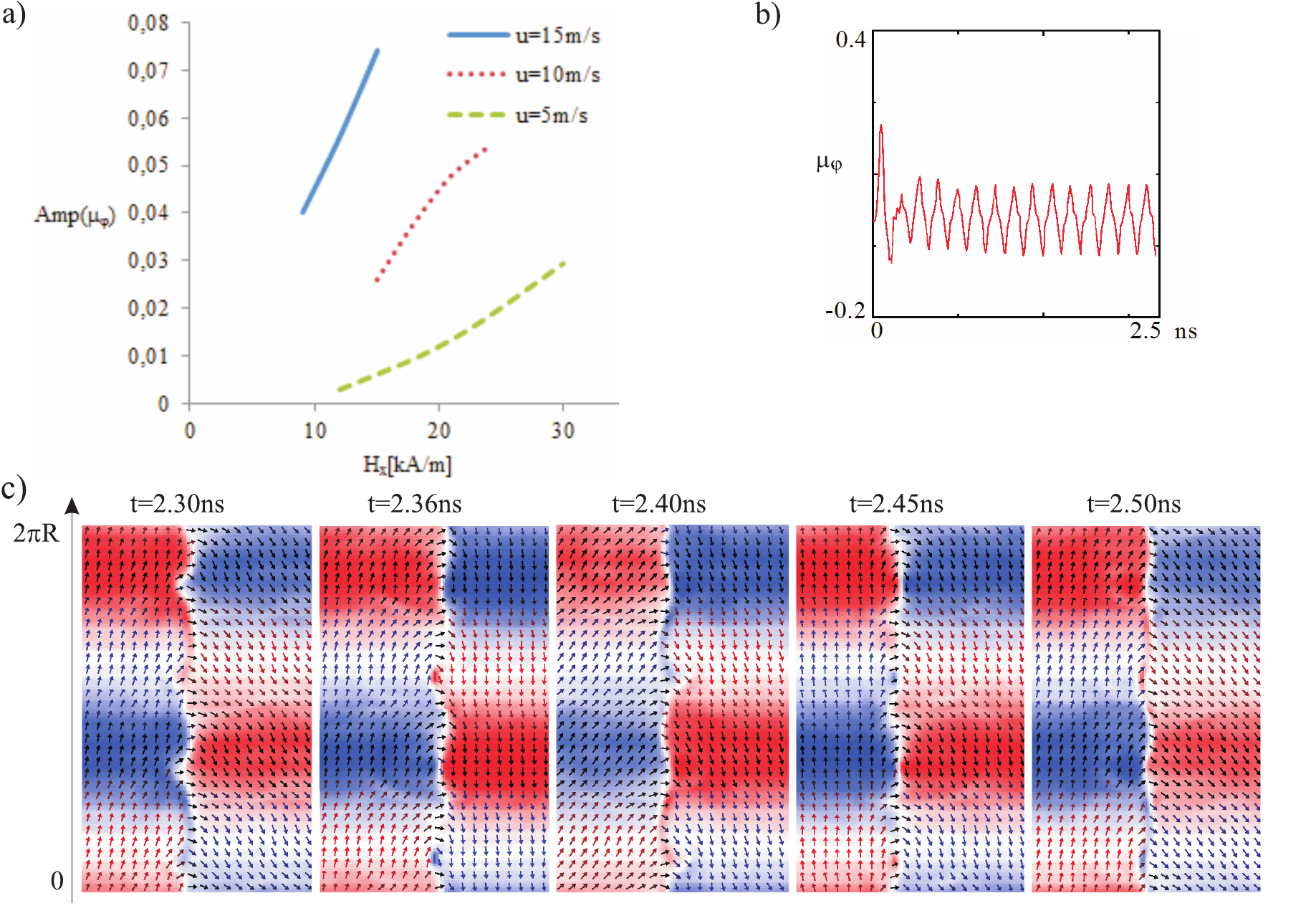}}}
\end{picture}
\end{center}
\caption{In (a)-(c), Simulations of 3$\mu$m-long tubular Co covering of the conducting nanowire (300nm diameter, 11nm thickness), 
relaxed from the uniform magnetization state ${\bf m}=(0,M,0)$. The amplitude of oscillations of the summary circumferential
magnetization of the nanotube with dependence on the longitudinal field for the current of the frequency $\nu=6$GHz (a).
Time dependence of the summary circumferential magnetization (b) and snapshots of the central 500nm-long area
of the nanotube (a single period of the DW evolution) (c) for $\nu=6$GHz and the scaled density u=15m/s
at the longitudinal field $H_{x}=15$kA/m.
The area colors and their intensity correspond to $y$-component of the magnetization.}
\end{figure*}

For different densities of the AC current along the tube/composite wire, we have established the time dependencies
 of the average [by means of (\ref{average_m})] circumferential component of the magnetization $\bar{m}_{\varphi}$ 
 and the time dependence of the induced electromotive force (EMF); (\ref{EMF}). The relative shift
 of the amplitude ${\rm Amp}[\epsilon]$ of EMF is equal to the MI ratio
\begin{eqnarray}
\Delta Z/Z(H_{x})=\frac{{\rm Amp}\left[\epsilon\right](H_{x})-{\rm Amp}\left[\epsilon\right](H_{ref})}{
{\rm Amp}\left[\epsilon\right](H_{ref})},
\label{mi-ratio}
\end{eqnarray}
where $H_{ref}$ is a reference value of the axial field. Here and below, ${\rm Amp}[\cdot]$ denotes 
 the amplitude of any oscillating quantity. Beyond the resonance
 range of the frequency, for $\nu_{c2}<\nu<\omega(0)/2\pi$, the MI ratio can be calculated as
\begin{eqnarray}
\Delta Z/Z(H_{x})\approx
\frac{{\rm Amp}[\bar{m}_{\varphi}](H_{x})-{\rm Amp}[\bar{m}_{\varphi}](H_{ref})}{
{\rm Amp}(\bar{H}_{\varphi})+{\rm Amp}[\bar{m}_{\varphi}](H_{ref})}
\end{eqnarray}
provided any higher-harmonic contributions to EMF are small. 
 In the majority of the MI studies, one takes the maximum field that allows for the impedance observation 
 to be a reference point $H_{ref}$. However, with regard to DW-based MI [$\Delta Z/Z(H_{x})$ is an increasing 
 function in a wide range of the field], it is convenient to take $H_{ref}=0$. The later choice 
 requires the EMF oscillations to be regular at zero field, which is shown below to be satisfied 
 for the nanotubes with a Neel DW while not with a cross-tie DW.

\subsection{Single domain wall in covering of long conducting nanowire}

The amplitude of the rescaled (via dividing by $M_{s}$) average (over the cross-section of the nanotube wall
 in $\varphi={\rm const}$ plane) circumferential magnetization; $\mu_{\varphi}\equiv\bar{m}_{\varphi}/M_{s}$
 (Figs. 3a-3c), and the MI ratio $\Delta Z/Z$ (Fig. 3d) are plotted with dependence on the axial field $H_{x}$
 for the Co nanotubes of the outer radius $R_{out}=$150nm, the length $L=3\mu$m, and the wall thickness of 11nm.
 The grid-discretization size in every simulation of the present work is 5nm. While simulating a nanostripe 
 with the periodic boundary condition is the alternative of simulating the genuine nanotube, we have chosen
 the later in order not to neglect possible effect of the surface curvature on the dynamics, (the static effects
 of the curvature are negligible in thin-walled nanotube \cite{ota13}).
 The simulations of the nanotube with a single DW have been performed for the current frequencies:
 0.3GHz, 3GHz, and 6GHz. It is comfortable to rescale the amplitude of the current density defining 
 $u\equiv\eta\cdot{\rm Amp}(j)$, where $\eta=P\cdot4.14\cdot10^{-11}$m$^{3}$/C, and taking the spin polarization 
 of cobalt P=0.4. 

We discuss details of the magnetization dynamics of the nanowire covering with a single Neel DW, focusing
 on the case of $\nu=6$GHz ($>\nu_{c2}$) and the current of $u=15$m/s, (the solid line in Figs. 3a,3b) for instance.
 Five ranges of the axial field are distinguished by different time courses of the average circumferential 
 magnetization $\mu_{\varphi}$. These specific field ranges are separated by its values
 $H_{x}=$12kA/m, 17kA/m, 70kA/m, 130kA/m. The regular oscillations of DW position (DW-based MI) correspond
 to the field area $H_{x}\in[0,17]$kA/m, which is  additionally divided into the regimes of the viscous
 ($H_{x}\in[0,12]$kA/m) and turbulent ($H_{x}\in(12,17]$kA/m) motions of the DW.
 The later type of the DW motion is accompanied by a sinusoidal shape of the DW line which circulates
 around the tube axis. During each AC cycle, there appear temporal germs of vortices
 in the DW structure, however, they do not transform into topologically-protected objects, (see snapshots in Fig. 3h).
 While uniform FMR governs the evolution of the magnetization (FMR-based MI) for $H_{x}\ge$130kA/m, 
 (the right hand side of the maximum of ${\rm Amp}[\mu_{\varphi}](H_{x})$ in Figs. 3a,3b), in the middle range
 of the field; $H_{x}\in(17,130)$kA/m, a structure of several domains appears and
 it coexists with the continuous rotation of the magnetization inside the domains. 
 In this middle-field range, a regime of chaotic oscillations of the average magnetization;
 $H_{x}\in(17,70)$kA/m is distinguished, (there is a gap in the solid line in Figs. 3a,3b since the oscillation
 amplitude is not determined). The irregular character of the oscillations is the effect of relatively large changes
 of the DW position mixed with the evolution within the domains. In the above context, some visible nonlinearity
 of the magnetization oscillations in the field regime $H_{x}\in(12,17]$kA/m (due to the turbulence of the DW motion)
 can be attributed to the continuous transition to chaos, that is usual for (Duffing-like) forced nonlinear
 oscillators, e.g. \cite{tay05}. In the second regime of the middle-field range; 
 $H_{x}\in[70,130)$kA/m, the domain structure is unstable; DWs are annihilated and created in each cycle
 of the AC current while the average-magnetization oscillations are regular, (the FMR dominates over the DW movement).
 The cyclic almost-complete annihilation of DWs results in a large amplitude of the magnetization oscillations, 
 (${\rm Amp}[\mu_{\varphi}]$ approaches 0.9).

The two basic mechanisms of MI of the thin-wall nanotube are explained by the following.
 In the low-field regime, MI of the system (with a single Neel DW) results from changes of the DW
 width $\Delta$ with changing the axial field, thus, changes of the maximum velocity of DW
 in a given alternating Oersted field, according to (\ref{q_H}). 
 In the high-field regime; the right-hand side of the maximum of ${\rm Amp}[\mu_{\varphi}](H_{x})$ in Figs. 3a,3b,
 MI is a consequence of the increase of the FMR frequency with increasing the axial field (see Fig. 2).
 At the field point of the low-frequency FMR; $H_{FMR}\approx H_{K}$, \cite{kra03},
 the normalized amplitude of the magnetization oscillations ${\rm Amp}[\mu_{\varphi}]$
 takes its maximum value (above 0.9). For $H_{x}>H_{FMR}$, increasing the field, one goes with
 the evolution conditions away from the resonance ones. Therefore, ${\rm Amp}[\mu_{\varphi}](H_{x})$ is decreasing 
 for $H_{x}>H_{FMR}$, however, the magnetization evolution remains a homogeneous rotation of the overall
 system around the long axis of the nanotube. 
 The transition from the (low-field) regime of DW-based MI to the (high-field) regime of FMR-based MI
 is accompanied by a qualitative change in the time course of the average circumferential magnetization. 
 In the low-field region, $\mu_{\varphi}(t)$ oscillates around some nonzero value (Figs. 3e,3f)
 while, in the high-field region, it oscillates around zero with a large amplitude (Fig. 3g). 
 
\begin{figure*} 
\unitlength 1mm
\begin{center} 
\begin{picture}(175,127)
\put(23,-7){\resizebox{150mm}{!}{\includegraphics{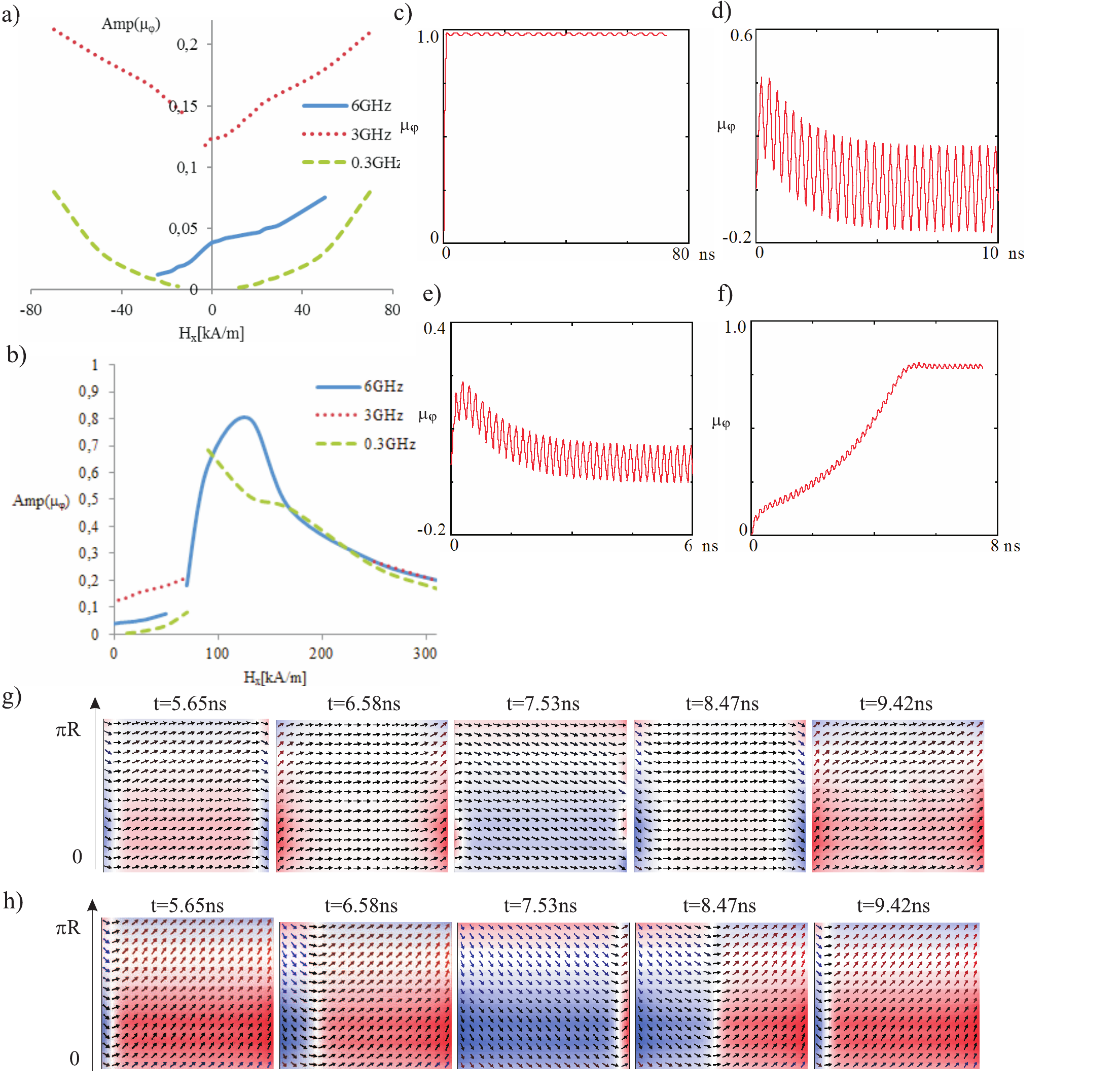}}}
\end{picture}
\end{center}
\caption{Simulations of 450nm-long tubular Co covering of the conducting nanowire (300nm diameter, 11nm thickness), relaxed from the uniform magnetization state ${\bf m}=(M,0,0)$ for the AC current of the scaled density u=5m/s.
The amplitude of oscillations of the summary circumferential magnetization of the nanotube with dependence on the longitudinal field (a),(b).
Time dependence of the summary circumferential magnetization for the longitudinal field $H_{x}=$24kA/m
for the current frequency $\nu=0.3$GHz (c), $\nu=3$GHz (d), $\nu=6$GHz (e), and for the field $H_{x}=$-24kA/m
and the frequency $\nu=6$GHz (f). 
The snapshots of whole-length of the nanotube (a single period of the DW evolution) for 
$\nu=0.3$GHZ and $H_{x}=170$kA/m (g), $\nu=0.3$GHZ and $H_{x}=90$kA/m (h).
The area colors and their intensity correspond to $y$-component of the magnetization.}
\end{figure*} 

Analyzing the DW evolution in the low-field regime, for the thin films of Co, we estimate the critical
 (circumferential) field of the Walker breakdown $H_{W}=\alpha M_{s}/2=35$kA/m. The current of the density $u=5$m/s in the 
 interior of the tube of 150nm radius induces the Oersted field $\bar{H}_{\varphi}=21$kA/m$=0.6H_{W}$ in the tube,
 while the current of $u=15$m/s induces the field well above the critical one; ${\rm Amp}(\bar{H}_{\varphi})=1.8H_{W}$.
 At the high current; $u=15$m/s, the maximum circular frequency of the hypothetical rotation of the DW magnetization
 takes the value of $\gamma\mu_{0}{\rm Amp}(\bar{H}_{\varphi})=2\pi\cdot0.8$GHz, 
 which is small compared to the circular frequency of the current $2\pi\cdot 6$GHz. Thus,
 despite overcoming the Walker breakdown field, the dynamical spin deviations inside the DW are not expected to be large.
 A consequence of the strong Oersted field (for $u=15$m/s) is a large shift
 of the maximum of the average magnetization with the current frequency (Fig. 3a). In particular,
 at $\nu=0.3$GHz (the dashed line in Fig. 3a), the position of the maximum of $\mu_{\varphi}(H_{x})$
 is close to $H_{x}=0$ while its value approaches 0.9, whereas, at $\nu=6$GHz, the maximum positon 
 is close to the anisotropy field $H_{K}$. The effect is reminiscent of shifting the maximum
 of the GMI ratio $\Delta Z/Z(H_{x})$ with changing the current frequency that has been reported 
 for large (close to the anisotropy field) amplitudes of the Oersted field \cite{pha08,vaz01}.
 Here, $u=15$m/s corresponds to the field amplitude ${\rm Amp}(\bar{H}_{\varphi})=63$kA/m$>0.3H_{K}$. 

At maximum, the MI ratio of the low-field regime (Fig. 3d) exceeds 100$\%$, thus, we can call
 the low-field effect GMI. For $u=15$m/s, the slope of the MI curve (the GMI sensitivity) reaches its maximum ($10\%$m/kA)
 at $H_{x}=10$kA/m. However, a direct comparison of the sensitivity value to experimental data 
 for other nanosystems is hindered by different choices of the reference field $H_{ref}$ 
 [in (\ref{mi-ratio})] and different operating ranges of the axial field. 
 A significant asymmetry of the MI curve in the vicinity of the zero field relates to the DW type that is 
 of the Neel structure. Its width increases or decreases with the longitudinal field depending whether the field 
 is oriented up or down the tube axis. For instance, for the current amplitude of $u=15$m/s, the MI is asymmetric
 in the range of $|H_{x}|<9$kA/m, while it is symmetric for $|H_{x}|\ge9$kA/m. The asymmetry vanishes 
 when the negative field is sufficiently high to reverse the magnetization inside DW. 

Consequences of the DW inertia (discussed in Section 2) are verified with simulations of the 
 nanotube with a single Neel DW under the current of a relatively-low density ($u=5$m/s), that produces 
 an Oersted field below the critical value of the Walker breakdown. In Fig. 4a, an irregularity of the magnetization
 oscillations is seen for zero field and circular frequency of the current ($2\pi\cdot0.3$GHz) well below
 the damping rate of the DW motion ($\Gamma=2\pi\cdot2.5$GHz). In a short period of time, the DW motion is cyclic
 while the effect of the inertia is accompanying the oscillations by large dynamical changes of the DW structure,
 (despite the Oersted field is low). The snapshots from a single cycle (Fig. 4c) show the creation and annihilation
 of the vortex-antivortex pairs inside DW. For the circular frequency of the current ($2\pi\cdot3$GHz)
 just above the damping rate $\Gamma$, the effect of inertia is not seen. The oscillations of the magnetization
 are stable (Fig. 4b), in phase with the driving current (sinusoidal), and not accompanied by changes
 of the DW structure.

Comparing the dynamics of the Neel DW under the AC current to the dynamics of the cross-tie DW (Figs. 5a-5c),
 (let us remind that both types of DW can coexist in the nanotube \cite{jan16}), we notice considerable differences.
 Since the cross-tie structure is not durable under the action of the Oersted field,
 the viscous motion of the cross-tie DWs (with a conserved DW structure)
 is not found for any value of the axial field. The current driven oscillations of the circumferential
 magnetization are irregular below a critical field of the transformation of the cross-tie DW to the Neel DW,
 thus, their amplitude is not determined.
 The instability of the cross-tie DW in the Oersted field results from movement of the vortices 
 and antivortices inside DW. Notice a difference from the widely-discussed motion of the vortex DW
 in the soft-magnetic nanostripes, which is stable in a wide range of low driving field (current).
 While, in the presence of the external field, a deformed vortex DW is stabilized by the stray field
 of the stripe edges \cite{cla08}, the cross-tie DW in the nanotube is not.
 Additionally, at the current frequency (of 6GHz) well above the relaxation rate (2.5GHz)
 and the field above the cross-tie-to-Neel threshold, the relaxation of the DW structure 
 from the cross-tie state can be incomplete, which explains a little-bit-larger dynamical
 distortions of DW in Fig. 5c than in Fig. 3h.
 
\subsection{Single domain wall in covering of short conducting nanowire or in empty nanotube}
  
\begin{figure*} 
\unitlength 1mm
\begin{center}
\begin{picture}(175,123)
\put(24,-6){\resizebox{130mm}{!}{\includegraphics{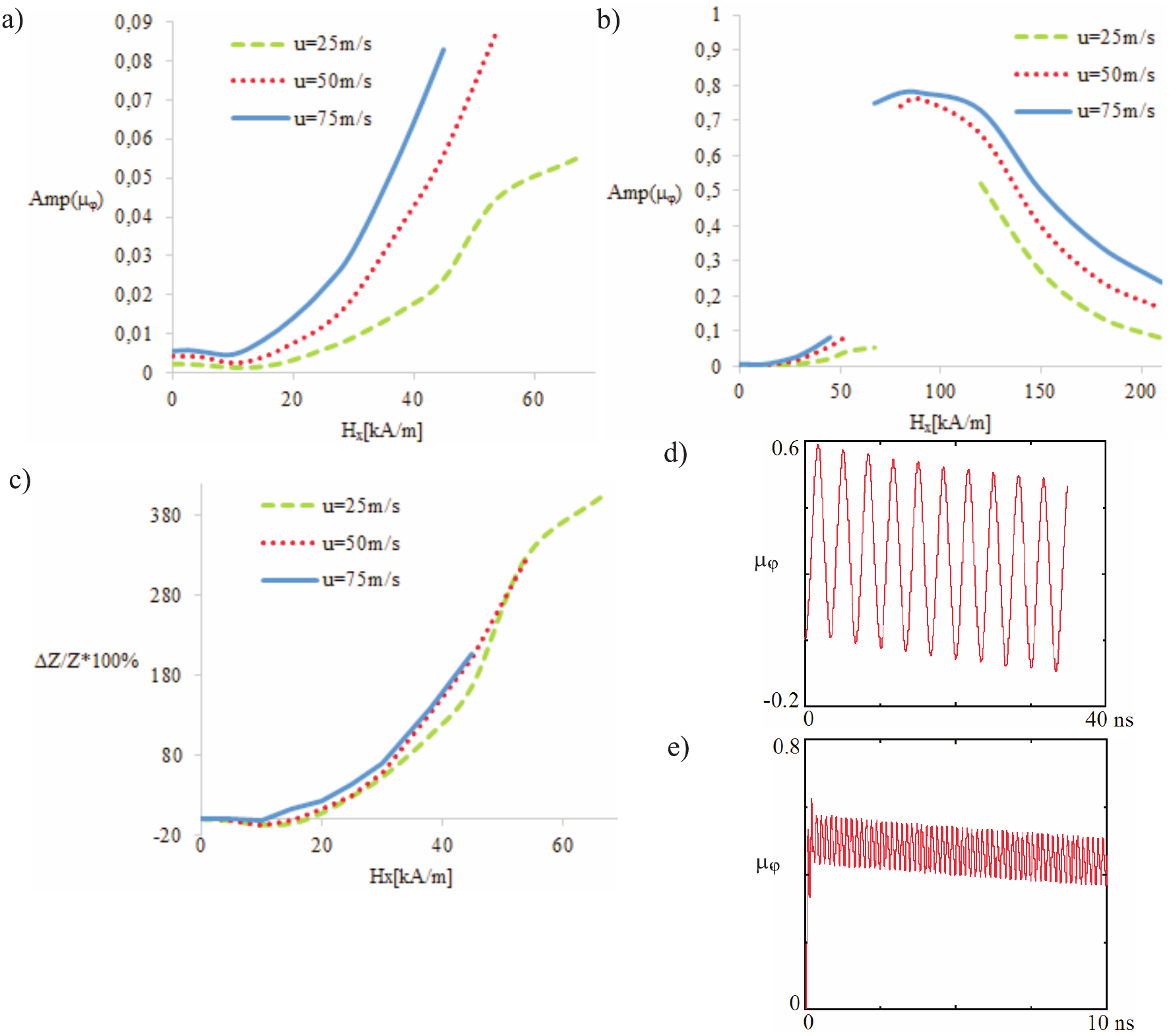}}}
\end{picture}
\end{center}
\caption{Simulations of an empty Co nanotube of 3$\mu$m length, 300nm diameter, 11nm thickness. 
The initial magnetization is relaxed from the uniform magnetization state ${\bf m}=(M; 0; 0)$,
the axial-anisotropy constant takes $K=-150$kJ/m$^{3}$. The amplitude of oscillations of the summary circumferential
magnetization of the nanotube (a),(b) and the relative
impedance (c) with dependence on the longitudinal field for the AC current of 6GHz.
Time dependence of the summary circumferential magnetization at the longitudinal field $H_{x}=40$kA/m,
for the current of the frequency of 0.3GHz and the current scaled density $u=25$m/s (d), 
for the frequency of 6GHz and $u=75$m/s (e). The area colors and their intensity correspond to $y$-component of the magnetization.}
\end{figure*}

In a short nanotube (a 450nm-long covering of a conducting nanowire), the effect of the tube ends on the motion
 of Neel DW reveals in a specific shape of the time dependence of the circumferential magnetization (Figs. 6c-6f).
 The sequence of points of the maximum slope of the oscillatory curves in Figs. 6d,6e; (the equilibrium position
 of the oscillations) exponentially tends to zero with time. This corresponds to a slow stabilization of the center
 of the DW-position oscillations at the tube center. The stabilization time decreases
 with increasing the axial field due to the field dependence of the DW width $\Delta$, thus, the field dependence
 of the strength of the DW interaction with the tube ends; [since the DW-binding force in (\ref{second-order-q})
 is dependent on $\Delta$].
 Notice that, in the 3$\mu$m-long covering of the conducting nanowire,
 the DW position oscillates under the AC current around any conserved point from a large sector of the tube, 
 thus, the effect of the tube ends is not seen in the studies of Subsection 3.1.

Studying dynamical characteristics of the 3$\mu$m-long empty nanotube with a single Neel DW
 under the AC current of a relatively large density $u=25\div75$m/s, we have found a similar to 
 the above described effect of the nanotube ends; the exponential slope of the envelope
 of the average magnetization $\mu_{\varphi}(t)$. Though, the effect (the slope in Figs. 7d,7e)
 is small compared to that in the 450nm-long nanotube (Figs. 6d,6e), it has not been observed 
 in the nanowire covering of the same sizes as the present nanotube. Since the oscillatory motion
 of DW (DW-based MI) is stable in the empty nanotube in a wider range of the axial field than
 in the nanowire covering, one is able to induce stronger interactions
 of DW with the tube ends in the former system (a larger DW width).

Notice that for the current density region of our interest ($u=25\div75$m/s), 
 the average Oersted field in the empty nanotube is well below
 the critical value of the Walker breakdown $H_{W}=35$kA/m, [e.g. at $u=25$m/s,
 one finds ${\rm Amp}(\bar{H}_{\varphi})=8$kA/m], while STT is relatively large
 ($\dot{q}_{j}/\dot{q}_{H}\sim 1$, see Appendix A). Estimated following \cite{bea08}, the current density
 of the threshold (a Walker-like breakdown) $u_{c}\equiv\eta j_{c}=H_{W}\gamma\mu_{0}\Delta/2\approx58$m/s
 belongs to the indicated range of our simulations while it increases with axial field due to the field dependence
 of the DW width $\Delta$. A noticeable feature of the MI curves of the empty nanotube (Figs. 7a,7c)
 while seen for the nanowire covering as well (Figs. 3c,3d)
 is a negative slope in the vicinity of the zero field. The decrease of the average magnetization $\mu_{\varphi}$
 and the MI ratio with increasing the axial field in the area of $H_{x}\in(0,10)$kA/m is in line with 
 (\ref{B_5}) provided the DW driving via STT dominates over the driving via the Oersted field;
 $\dot{q}_{j}>\dot{q}_{H}$. This is because $\dot{q}_{j}$ is independent of the field, unlike 
 $\dot{q}_{H}\propto\Delta(H_{x})$.

The case of the short (450nm-long) nanowire covering at high frequency of the current (6GHz) and negative axial field
 $H_{x}\in[-24,-9]$kA/m is specific in terms of the course of the envelope of the average magnetization
 $\mu_{\varphi}(t)$, (see Fig. 6f). 
 The envelope tends with time to a high-lying ($|\mu_{\varphi}|\sim 0.7\div 0.8$) plateau,
 (provided the axial field is not applied prior to the current). Because of incomplete relaxation
 (the DW relaxation rate is small compared to the current frequency), the axial component of the DW magnetization
 remains opposite to the field while its vector deviates by an angle from the tube axis.
 This deviation seems to change the character of the long-distance interaction of DW with the tube ends
 from the repulsion to the attraction. The plateau of $\mu_{\varphi}$ corresponds to an oscillatory motion
 of DW in the vicinity of the tube end.
 
Another effect of the tube ends is a resonance increase of the amplitude of the DW-position oscillations
 in the frequency region $\nu_{c1}<\nu<\nu_{c2}$. The parameter of the magnetic hardness of the tube $\delta$
 [that relates to the ratio of thickness of the tube wall to its diameter and determines the binding-force
 coefficient $\omega_{0}$ of (\ref{second-order-q}); $\omega_{0}\propto(\delta/L)^{1/2}$]
 is sufficiently large that the driven oscillations of the DW position in the 450nm-long
 nanotube are not overdamped, thus, $\Gamma<\sqrt{2}\omega_{0}$. With the above requirement, we estimate 
 the minimal frequency range of the resonance for the Co nanotube via
 $\nu_{c2}>(\sqrt{3/4}+1/2)\Gamma/2\pi=3.4$GHz and $\nu_{c1}<(\sqrt{3/4}-1/2)\Gamma/2\pi=0.37$GHz.
 For the AC amplitude $u=5$m/s and zero axial field, the analysis of
 the $\nu=6GHz$ data in Fig. 3c and Fig. 6a as well as the $\nu=3GHz$ data in Fig. 4b and Fig. 6a
 shows faster increase of the amplitude of the magnetization oscillations
 with decreasing the tube length for 3GHz than for 6GHz.
 We explain that by belonging of the frequency 3GHz to the resonance area (3GHz$<\nu_{c2}<$6GHz).
 For the frequency of 0.3GHz, at $u=5$m/s, we observe the resonance-driven transition to the single-domain state
 via the DW collision with one of the tube ends; $\mu_{\varphi}$ approaches its maximum value 1.0. 
 In this case, the single-domain circumferentially-magnetized state stabilizes
 in the short nanotube for any value of the longitudinal field up to $70$kA/m.
 Observed in the field range $|H_{x}|\in[12,70]$kA/m, small-amplitude
 oscillatory dynamics of the circumferential magnetization (Fig. 6c) results from a uniform rotation
 of the magnetization far from FMR, however.
 
Comparing the low-frequency (0.3GHz) curves of $\mu_{\varphi}(H_{x})$, for the long and short coverings
 of the nanowire, in the high-field (FMR-based MI) range, one sees the positions of their maxima
 for the short nanotube (Fig. 6b) to be shifted upward the field axis 
 relative to the maximum for the long nanotube
 (Fig. 3a). According to the theory, the position of the maximum corresponds to the 
 critical field of the low-frequency FMR which is evaluated in \cite{kra03}
 to correspond to the anisotropy field $H_{FMR}\approx H_{K}=2|K|/\mu_{0}M_{s}$, based on (\ref{FMR}). 
 For our nanotube of Co, we have $H_{K}=170$kA/m, 
 which is the position of the inflection point of the dashed curve in Fig. 6b (for $\nu=0.3$GHz)
 and the maximum point of the dotted line (for $\nu=3$GHz). The relevant dynamics visualized in the snapshots
 of Fig. 6g is usual for FMR. The absolute maximum of the dashed line in Fig. 6b
 (at $H_{x}\approx90$kA/m) corresponds to the snapshots of Fig. 6h, which show the dynamics to 
 be an oscillatory motion of the DW between both ends of the tube, in the presence of large deviation of the 
 domain magnetization from the main axis of the tube. Such an ordering of the domains prevents DW from
 the annihilation at the nanotube end. 
 
Significant values of the GMI ratio found for the 3$\mu$m-long empty nanotube are to be stressed (Fig. 7c). 
 In the low-field regime (DW-based MI) at $\nu=6$GHz and $u=25\div50$m/s, this ratio exceeds 200$\%$
 in the vicinity of the field $H_{x}=50$kA/m. However, the MI curves for different current densities overlap
 in a wide range of the field. 
 
\subsection{Several domain walls in a nanotube}

\begin{figure*} 
\unitlength 1mm
\begin{center}
\begin{picture}(175,189)
\put(24,-6){\resizebox{130mm}{!}{\includegraphics{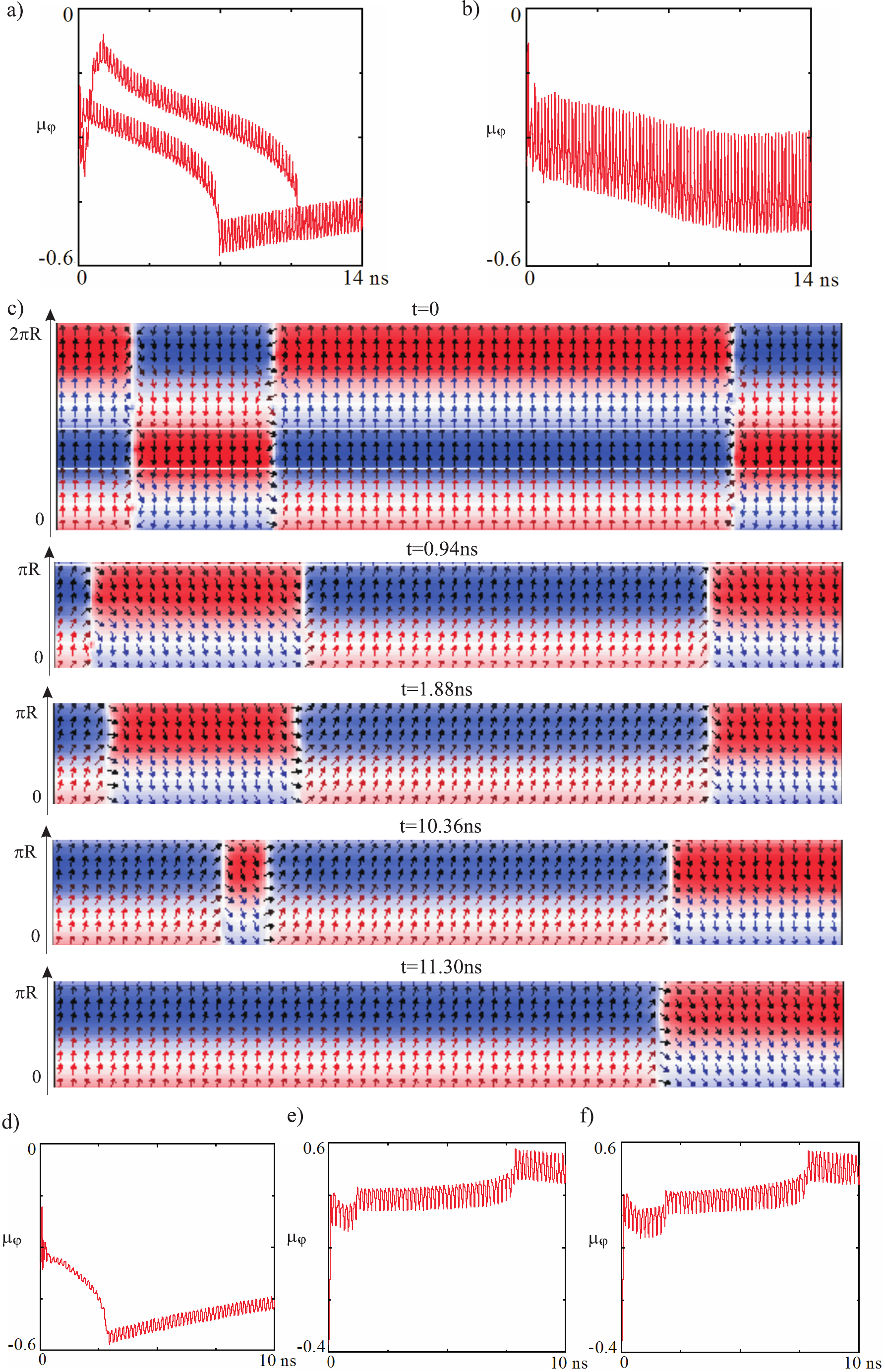}}}
\end{picture}
\end{center}
\caption{Simulations of a tubular Co covering of the conducting nanowire (a)-(c) and an empty nanotube
(d)-(f) of 3$\mu$m length, 300nm diameter, 11nm thickness, with the AC current of 6GHz. 
The initial magnetization is relaxed from a disordered state, the axial-anisotropy constant takes $K=-150$kJ/m$^{3}$.
The time dependence of the summary circumferential magnetization 
with the scaled current density u=5m/s at the longitudinal field $H_{x}=$30kA/m (upper curve of a) 
and with u=15m/s and $H_{x}=$15kA/m (b), u=25m/s and $H_{x}=$40kA/m (d), u=75m/s and $H_{x}=$40kA/m (e), 
u=75m/s and $H_{x}=$40kA/m and without STT (f).
The snapshots of whole the nanotube length for u=5m/s and $H_{x}=$30kA/m (c) correspond to the upper curve
of a). Additionally, the time course of the circumferential magnetization is plotted 
for the system relaxed before the current switching on with u=5m/s and $H_{x}=$30kA/m (lower curve of a).
The area colors and their intensity correspond to $y$-component of the magnetization.}
\end{figure*} 

Finally, we have simulated systems of several DWs in the tubular Co covering
 of a conducting nanowire and in the empty Co nanotube. A state of three cross-tie DWs 
 in the 3$\mu$m-long nanotube have been relaxed from a disordered state \cite{jan16}.
 From Subsection 3.1, the AC-driven motion of the cross-tie DWs 
 is known to induce irregular (chaotic) oscillations of EMF at very low values 
 of the longitudinal field, (below the cross-tie-to-Neel threshold). Thus, the field regime
 above the threshold value $|H_{x}|\ge9$kA/m is of our interest.   
 Our purpose was to evaluate the influence of the mutual interactions of DWs on MI.
 If there were no interactions, the increase of the number of DWs would be expected to result
 in the multiplication of the amplitude of oscillations of overall (average) circumferential 
 magnetization, according to (\ref{average_m}). Thus, increasing the DW number is considered
 to be the way to enhance the MI efficiency (the MI ratio and its sensitivity). 

Unfortunately, in the simulated systems, the nanowire covering and the empty nanotube with three DWs
 (of the cross-tie type), we have not managed to establish the long-term-durable oscillations
 of the circumferential magnetization. Their amplitude is constant within periods of dozen oscillations
 while it jumps upon these periods due to interactions of the DWs
 with the tube ends and/or with each other (Figs. 8a,8b,8d-8f).
 During the periods of the oscillation stability, the amplitude of the magnetization oscillations 
 is being smaller than triple of amplitude of the oscillations in the relevant single DW system.
 For example, from Fig. 8b, that is performed for the nanowire covering
 with three DWs, with the AC current of $u=15$m/s, in the field $H_{x}=15$kA/m, we find the maximum value
 of the amplitude of the average magnetization ${\rm Amp}(\mu_{\varphi})\approx 0.12$, which is about
 3/2 of value found for the similar nanowire covering with a single DW under the similar field
 and current (Figs. 3c,5a). Thus, induced EMF does not scale linearly with the number
 of DWs, in contrast to the expectation from (\ref{B_4}), \cite{atk98,mac96}. 
 We explain this by mutual interactions of DWs and their interactions with the tube ends which,
 therefore, cannot be neglected in a realistic model of the nanotube dynamics. 
 The long-range dipolar interactions destabilize the oscillating system, while, the short-range
 exchange interactions are responsible for the annihilation of the pair of colliding DWs magnetized in the same
 direction (in the tubular covering of the nanowire as well as in the empty nanotube),
 which is seen e.g. from the snapshots in Fig. 8c. Unexpectedly, studying the corresponding
 time dependence of the circumferential magnetization (upper curve in Fig. 8a), one finds the annihilation
 of the DW pair not to result in any decrease of the amplitude of the magnetization oscillations.
 This confirms the effect of the DW interactions on the magnetization evolution to be strong.

Note that the DW annihilation requires their interaction to be attractive and, in the system under 
 an axial field above the threshold value (of the cross-tie-to-Neel transition inside DWs), they are so
 since all DWs are magnetized in the same direction \cite{jan13}. However, in order to exclude
 any delayed effects of incomplete relaxation of the DW structure on their collision,
 we have simulated the system, switching the AC current on simultaneously with the axial field,
 as well as switching the current on upon the axial field (upon Neel DWs had relaxed).   
 The long-term dynamics appears to be independent of the current-switching conditions except
 certain consequences of different positions of DWs at the final moment of the DW-structure
 relaxation (compare the lower and upper curves in Fig. 8a). 
 A short-term effect of a delayed relaxation of one DW due to the tube-end proximity
 is seen with the presence of a vortex in the structure of that DW; 
 on the left-hand side of the snapshot $t=0.94$ns in Fig. 8c.
      
Simulating the driven dynamics of the empty nanotube with three DWs, we have observed 
 a qualitative change of the time-course of the magnetization with changing the amplitude 
 of the current density, (compare Figs. 8d,8e). Increasing the current, 
 the moment of the DW annihilation [a global extremum of $\mu_{\varphi}(t)$]
 shifts upward the time axis, while the circumferential
 magnetization changes its sign at some threshold value of the current amplitude.
 In particular, in the field of $H_{x}=40$kA/m, the left-hand-side pair of the DWs collide and annihilate
 at low current ($u=25$m/s) at the time $t\approx 3$ns (Fig. 8d) while, under the current of $u=75$m/s,
 the right-hand-side pair collide at the time $t\approx 8$ns (Fig. 8e). That dynamics change with the current 
 amplitude seems to result from the DW interaction with the tube ends. In order to confirm the importance
 of those interactions, we have performed the simulations with the same field ($H_{x}=40$kA/m) and the current
 amplitude $u=75$m/s while omitting STT (taking $\eta=0$), which is known not to affect
 the mutual DW interactions (Fig. 8f). Comparing Fig. 8e and Fig. 8f,
 we notice the second magnetization jump in both plots, that corresponds to the DW-pair annihilation, 
 to be placed at the same point of time ($t\approx 8$ns) and the amplitude of the magnetization oscillations
 not to be noticeably influenced by STT in the strong axial field applied, (we mention that the DW-velocity
 component $\dot{q}_{j}$ is independent of the axial field, unlike $\dot{q}_{H}\propto\Delta$).
 However, STT influences the position of the DW system relative to the tube. A shift in time of the 
 first jump in Fig. 8f relative to Fig. 8e confirms the presence of the DW collision with
 the tube end. With some choices of the current and field parameters, we have found two such jumps
 in $\mu_{\varphi}(t)$, before another jump due to the DW-pair annihilation.
  
\section{Conclusions}

We have studied MI of the ferromagnetic thin-wall nanotubes with the micromagnetic simulations, 
 focusing on the low-field regime of the magnetization dynamics, that 
 relates to the oscillations of the position of (single) DW. Unlike in $\mu$m-sized systems,
 the DW motion along the nanotube can be isolated from other-type magnetization oscillations
 for up to the GHz-frequency range. The ratio of DW-based MI reaches 100$\%$
 for the tubular Co covering of a conducting nanowire and 200$\%$ for empty Co nanotube.
 Within the field range of DW-based MI, the operating range of a significant GMI sensitivity
 (a differential of the MI ratio $\Delta Z/Z$ over the field $H_{x}$) is relatively wide.
 For our Co nanotubes it is up to several tens of kA/m, however, this is mainly due to the choice
 of the nanotube material; [because of the large saturation magnetization of cobalt, a very strong 
 anisotropy field $H_{K}$ is required in order to counteract the magnetostatic field, which influences 
 the position of FMR peak in $\Delta Z/Z(H_{x})$].
 Since linear magnetic field dependence of the output signal is desirable for sensor applications \cite{zhu13},
 for a given operating range of the field, a relatively low GMI sensitivity of the nanotube 
 (up to 10$\%$m/kA) is not any disadvantage. 
 In connection to the sensing potential of nanotube GMI, let us notice that the magnetic fields produced
 in nanoscale devices can be very high locally, for instance, the Oersted fields due to high-density currents.
 
The (magnetostatic) effect of the tube length on GMI (previously studied with relevance to the microtubes;
 thick coverings of conducting non-magnetic microwires of the diameter of 100$\mu$m, \cite{elk16})
 is important with respect to optimizing GMI of the nanotubes.
 Similar to the microtube, changing the nanotube length results in shifting the position of the  
 (originating from FMR) peak of the MI ratio $\Delta Z/Z(H_{x})$ while the effect is very weak.
 However, the nanotube length influences strongly the low-field (DW-based) GMI, in particular,
 via widening the operating range of the field with shortening the nanotube. 
 That desirable effect (a magnetic hardening within the domains)
 is accompanied by a resonance increase of the amplitude of the DW-position
 oscillations under AC current. Moreover, regular oscillations of the magnetization
 of short nanotubes are present in the transient area of the field, when the FMR-like magnetization
 dynamics is correlated with large-amplitude oscillations of the DW position.
 Notice that the resonant motion of DW in the nanotube is very different from its counterpart 
 in microwires/microtubes because of different origins of the binding potential \cite{zim15a}.
 In the later systems, the dominant restoring force results from a dynamical flexure of DW
 (due to pinning centers in the material) \cite{kle17}.  
 
We have verified the applicability of a simple description
 of DW-based MI of the circumferentially-magnetized many-domain tube that neglects the 
 DW interactions to the nanosystems \cite{atk98,mac96}.
 The interactions of the DWs with each other and with the nanotube ends cannot be neglected 
 at the nanoscale, especially, in the presence of a strong longitudinal field that stabilizes
 the Neel-DW structure. That field magnetizes all the DWs in the same direction, thus, causing their 
 interactions to be attractive and allowing for their annihilations during collisions. 
 The most stable oscillations of the circumferential magnetization 
 are found in the nanotubes that contain a single DW. 
 As discussed in Ref. \cite{jan16}, such a domain structure can be created via the tube relaxation
 from a state of the uniform magnetization.
   
Note that, focusing on the DW-based effect, we have not examined GMI in the so called "single-peak" regime
 of very-high current densities, whose ratio $\Delta Z/Z(H_{x})$ takes
 its maximum at $H_{x}=0$, \cite{vaz01}. Such high currents (that induce the Oersted fields above the 
 the circumferential coercivity \cite{pha08}) would result in producing a large amount of the Joule heat, however.
 With the formula for the slope of the time-dependence of temperature ${\rm d}T/{\rm d}t=j^{2}/\rho_{Co}C_{Co}\sigma_{Co}$,
 with the mass density of cobalt $\rho_{Co}=8900$kg/m$^{3}$, the specific heat $C_{Co}=420$J/kg$\cdot$K,
 and the electrical conductivity $\sigma_{Co}=(6.2\cdot10^{-8}\Omega\cdot{\rm m})^{-1}$, \cite{fan11},
 for DC current of the density $j=10^{12}$A/m$^{2}$, we estimate ${\rm d}T/{\rm d}t=17$K/ns. Including the sinusoidal
 time dependence of the current density, with ${\rm Amp}(j)=10^{12}$A/m$^{2}$; (thus, $u=17$m/s, 
 which is about maximum density of current in filled nanotubes we simulated), and averaging its square over the single period,
 we obtain $\widetilde{{\rm d}T/{\rm d}t}=8.5$K/ns to be an upper bond on the heating pace in the nanowire coverings,
 provided Joule heating in the tube-filling nanowire is not faster. It is an acceptable value, especially, 
 for the systems of cobalt whose Curie temperature is very high. However, the empty nanotubes are operated
 with several-times-higher current amplitudes, on the limit of the overheating.  
 
\section*{Acknowledgement}

Calculations have been carried out using resources provided by Wroclaw Centre for Networking and Supercomputing
 (http://wcss.pl), grant No. 450.
 
\section*{Appendix A. AC-driven motion of domain wall}
 
Following \cite{sch74,thi05,bea07}, the DW motion in the 
 stripe-like system is governed by the dynamical equations of its position $q$
 and the angle $\phi$, that measures the deviation of the DW magnetization
 from the plane of the ferromagnetic film about the easy axis
\begin{eqnarray}
\dot{\phi}+\frac{\alpha}{\Delta}\dot{q}&=&\gamma\mu_{0}\bar{H}_{\varphi}+\frac{\beta\eta}{\Delta}j,
\nonumber\\
\dot{q}/\Delta-\alpha\dot{\phi}&=&\frac{\gamma\mu_{0}H_{M}}{2}\sin(2\phi)
\label{q-phi}\\
&&+\gamma\mu_{0}H_{x}\sin(\phi)+\frac{\eta}{\Delta}j.
\nonumber
\end{eqnarray}
Here, $H_{M}$ is proportional to the difference of radial and axial anisotropy constants;
 $H_{M}\equiv M_{s}-2|K|/\mu_{0}M_{s}\approx M_{s}$ by analogy to the infinite planar
 ferromagnet \cite{sch74}. The analogy is valid since the tube length is much larger than
 the thickness of its wall. The field $\bar{H}_{\varphi}$ denotes the averaged
 over the tube cross-section Oersted field, and $\Delta$ denotes the DW width at rest
 that can be expressed with the magnetostatic and longitudinal-anisotropy exchange lengths 
 $l_{ms}\equiv\sqrt{2A_{ex}/\mu_{0}M^{2}}$ and $l_{K}\equiv\sqrt{A_{ex}/K}$, respectively;
 $\Delta=(l_{K}^{-2}+m_{\rho}^{2}M^{-2}l_{ms}^{-2})^{-1/2}\approx l_{K}$ for $H_{x}=0$.
 Here, $m_{\rho}$ denotes the radial component
 of the magnetization of DW, which is equal to zero in the case of Neel DW. 
 
The field and current dependencies of the DW velocity are well established for the case
 of the stationary motion (in a driving field $\bar{H}_{\varphi}={\rm const}$)
 via solving the above dynamical equations. We adapt such solutions to the case 
 of time dependent (oscillating) field $\bar{H}_{\varphi}(t)$ an current $j(t)$.
 Denoting in $q=q_{H}+q_{j}$ the Oersted-field and STT contributions 
 with $q_{H}$ and $q_{j}$, respectively, we use the classic Walker relation
\begin{eqnarray}
\dot{q}_{H}=\left\{\begin{array}{cc}
\frac{\gamma\mu_{0}\Delta\bar{H}_{\varphi}}{\alpha} & |\bar{H}_{\varphi}|<H_{W},\\
\frac{\alpha\gamma\mu_{0}\Delta\bar{H}_{\varphi}}{1+\alpha^{2}} & |\bar{H}_{\varphi}|\gg H_{W},
\end{array}
\right.
\label{q_H}
\end{eqnarray}
where $H_{W}=\alpha M_{s}/2$ denotes a critical field, and 
\begin{eqnarray}
\dot{q}_{j}=\left\{\begin{array}{cc}
\frac{\beta}{\alpha}\eta j & |j|<j_{c},\\
\eta j & |j|\gg j_{c},
\end{array}
\right.
\label{q_j}
\end{eqnarray}
where $j_{c}$ is the current counterpart of the critical field of the Walker breakdown.
 In a weak-driving case, (${\rm max}|\bar{H}_{\varphi}(t)|<H_{W}$ and ${\rm max}|j(t)|<j_{c}$),
 the angle $|\phi|\ll\pi/2$ and the position of DW oscillates in phase with the 
 the current. 

The ratio $\dot{q}_{j}/\dot{q}_{H}$ is a measure of efficiency of STT in driving
 the DW motion compared to the driving by the Oersted field. 
 In the case of the system simulated in Section 3, for the empty nanotubes,
 this ratio is about 10 times larger than for the nanowire coverings.  

The magnetostatic effect of the nanotube ends on the oscillations of the DW position
 can be significant in sufficiently-short nanotubes or under sufficiently-high AC current. 
 Including it, we restrict our description to tubes with a single DW and we modify the circumferential
 magnetic field by the following
 $\bar{H}_{\varphi}\to\bar{H}_{\varphi}+(1-2q/L)\delta M_{s}$. Here, $L$ denotes the tube length;
 ($q\in[0,L]$), while $\delta\in(0,1)$ depends on the thickness of the tube wall, ($\delta\to0$ in the limit
 of zero thickness, and $\delta\to1$ in the limit of the wire).
 The dynamical equations (\ref{q-phi}) can be decoupled into
\begin{eqnarray}
\ddot{q}+\frac{\alpha\gamma\mu_{0}}{1+\alpha^{2}}\left[H_{M}\cos(2\phi)+\frac{2M_{s}\delta\Delta}{L}\right]\dot{q}
\nonumber\\
+\frac{2\gamma^{2}\mu_{0}^{2}H_{M}M_{s}\delta\Delta}{(1+\alpha^{2})L}\cos(2\phi)\left(q-\frac{L}{2}\right)=
\nonumber\\
=\frac{\alpha}{1+\alpha^{2}}\left[\gamma\mu_{0}\Delta\dot{\bar{H}}_{\varphi}+\left(\frac{1}{\alpha}+\beta\right)\eta\dot{j}\right]
\nonumber\\
+\frac{\gamma\mu_{0}H_{M}}{1+\alpha^{2}}\left(\gamma\mu_{0}\Delta\bar{H}_{\varphi}+\beta\eta j\right)\cos(2\phi),
\label{second-order-q}\\
\ddot{\phi}+\frac{\alpha\gamma\mu_{0}}{1+\alpha^{2}}\left[H_{M}\cos(2\phi)+\frac{2M_{s}\delta\Delta}{L}\right]\dot{\phi}
\nonumber\\
+\frac{2\gamma^{2}\mu_{0}^{2}H_{M}M_{s}\delta\Delta}{(1+\alpha^{2})L}\frac{\sin(2\phi)}{2}=
\nonumber\\
=\frac{\alpha}{(1+\alpha^{2})\Delta}\left[\gamma\mu_{0}\Delta\dot{\bar{H}}_{\varphi}+\left(\frac{\beta}{\alpha}+1\right)\eta\dot{j}\right]
\nonumber\\
+\frac{2\gamma\mu_{0}M_{s}\delta\eta}{(1+\alpha^{2})L}j.
\end{eqnarray}
Assuming the amplitude of the oscillations of $\phi$ not to be large, we substitute $\phi=0$ into
 (\ref{second-order-q}), thus, finding the DW position in a finite-length nanotube 
 to be governed by the equation of the damped-driven oscillator.

\section*{Appendix B. domain-wall-based magnetoimpedance}

\begin{figure} 
\unitlength 1mm
\begin{center}
\begin{picture}(78,57)
\put(-4,-2){\resizebox{86mm}{!}{\includegraphics{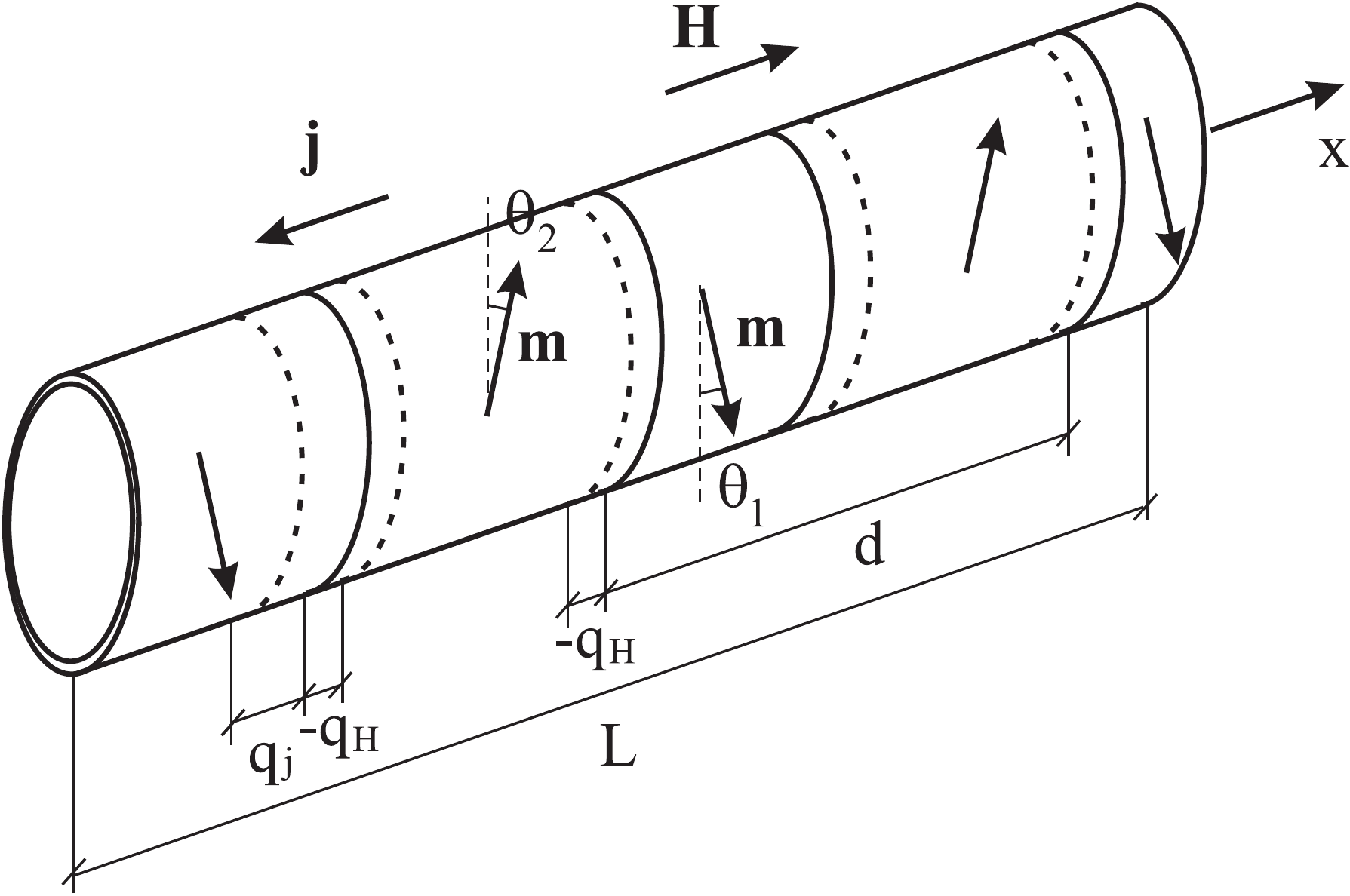}}}
\end{picture}
\end{center}
\caption{Scheme of the domain structure of the nanotube, (d is a "period" of the domain structure).}
\end{figure} 

In a nanotube-containing circuit, EMF is directly related to the time derivative
 of the circumferential magnetization of the tube \cite{vel94},
\begin{eqnarray}
\epsilon
=\int_{R_{in}}^{R_{out}}\int_{0}^{L}\mu_{0}\left(\dot{m}_{\varphi}
+\dot{H}_{\varphi}\right){\rm d}x{\rm d}\rho.
\label{EMF}
\end{eqnarray} 
Following a macroscopic model of MI due to the DW motion \cite{mac96}, the circumferential component
 of the overall magnetization of the tube-wall cross-section depends on a difference in length of the 
 clockwise- and counter-clockwise-magnetized domains
 and on the angles of deviation of the magnetization of both kinds of the domains from the circumferential direction
 of the tube $\theta_{1}$, $\theta_{2}$, (Fig. 9). Those angles are largely affected by the axial
 field applied. The cross-section average of the circumferential magnetization takes the form
\begin{eqnarray}
\bar{m}_{\varphi}\equiv\frac{1}{L(R_{out}-R_{in})}\int_{R_{in}}^{R_{out}}\int_{0}^{L}m_{\varphi}{\rm d}x{\rm d}\rho
\nonumber\\
=M_{s}\left[\frac{1}{2}\left(\cos\theta_{1}-\cos\theta_{2}\right)
\right.\label{average_m}\\\left.
+\left(\frac{Nq_{H}}{L}+\frac{\tau q_{j}}{L}\right)\left(\cos\theta_{1}+\cos\theta_{2}\right)\right],
\nonumber
\end{eqnarray}
where $N$ denotes the number of DWs in the tube, while $\tau=1$ for odd $N$ or $\tau=0$ for even $N$. 
 Hence, EMF is equal to
\begin{eqnarray}
\epsilon=\mu_{0}M_{s}(R_{out}-R_{in})\left(N\dot{q}_{H}+\tau\dot{q}_{j}\right)\nonumber\\
\times
\left(\cos\theta_{1}+\cos\theta_{2}\right)+\mu_{0}L(R_{out}-R_{in})\dot{\bar{H}}_{\varphi}.
\label{B_4}
\end{eqnarray}

One evaluates the angles $\theta_{1}$, $\theta_{2}$ via minimizing the relevant part of the magnetic energy
 of the tube $E=E_{an}+E_{Zx}+E_{Z\varphi}$, where
 $E_{an}$ denotes the anisotropy energy, while the Zeeman energy has been divided
 into the contribution from an axial magnetic field $E_{Zx}$
 and the one from a circumferential (Oersted) field $E_{Z\varphi}$
\begin{eqnarray}
E_{an}=-K\left[\lambda\sin^{2}\theta_{1}+(1-\lambda)\sin^{2}\theta_{2}\right],
\nonumber\\
E_{Zx}=-\mu_{0}M_{s}H_{x}\left[\lambda\sin\theta_{1}+(1-\lambda)\sin\theta_{2}\right],
\\
E_{Z\varphi}=-\mu_{0}M_{s}H_{\varphi}\left[\lambda\cos\theta_{1}-(1-\lambda)\cos\theta_{2}\right].
\nonumber
\end{eqnarray}
Here $\lambda=1/2+(Nq_{H}+\tau q_{j})/L$. The extremum conditions $\partial E/\partial\theta_{1}=0$,
 $\partial E/\partial\theta_{2}=0$, with $|H_{\varphi}|\ll H_{K}$, lead to $\sin\theta_{1}=\sin\theta_{2}=H_{x}/H_{K}$.
 Therefore, 
\begin{eqnarray}
\epsilon=2\mu_{0}M_{s}(R_{out}-R_{in})\left(N\dot{q}_{H}+\tau\dot{q}_{j}\right)
\nonumber\\
\times
\sqrt{1-(H_{x}/H_{K})^{2}}
+\mu_{0}L(R_{out}-R_{in})\dot{\bar{H}}_{\varphi}.
\label{B_5}
\end{eqnarray}


\section*{References}

\begin{thebibliography}{99}  
\bibitem{kno03}M Knobel, M Vazquez, L Kraus, Giant Magnetoimpedance, in: K.H.J. Buschow (Eds.), Handbook of Magnetic Materials, 
(Elsevier 2003)
\bibitem{pha08}M.-H. Phan, H.-X. Peng, Giant magnetoimpedance materials:
Fundamentals and applications, Prog. Mater. Sci. 53 (2008) 323.
\bibitem{zhu09}V. Zhukova, M. Ipatov, A. Zhukov, Thin Magnetically Soft Wires for Magnetic Microsensors,
Sensors 9 (2009) 9216. 
\bibitem{zhu16}V. Zhukova, Soft Magnetic Wires for Sensor Applications, in: A. Zhukov (Eds.), Novel Functional Magnetic Materials, (Springer 2016)
\bibitem{kur11}G. V. Kurlyandskaya, N. G. Bebenin, V. O. Vas'kovsky, Giant magnetic impedance
of wires with a thin magnetic coating, Phys. Metals Metallogr. 111 (2011) 133.
\bibitem{zhu18}V. Zhukova, J. M. Blanco, A. Chizhik, M. Ipatov, A. Zhukov, AC-current-induced magnetization switching
in amorphous microwires, Front. Phys. 13 (2018) 137501.
\bibitem{atk98}D. Atkinson, P. T. Squire, Phenonemological model for magnetoimpedance
in soft ferromagnets, J. Appl. Phys. 83 (1998) 6569.
\bibitem{zim15}J. Ziman, V. Suhajova, M. Kladivova, Effect of domain structure on the impedance
of ferromagnetic wire with circumferential anisotropy, Sensors and Actuators A 223 (2015) 134.
\bibitem{val02}R. Valenzuela, The analysis of magnetoimpedance by equivalent circuits, J. Magn. Magn. Mat.
249 (2002) 300.
\bibitem{pan04}L. V. Panina, D. P. Makhnovskiy, K. Mohri, Magnetoimpedance in amorphous wires and multifunctional
applications:from sensors to tunable artificial microwave materials, J. Magn. Magn. Mater. 272-276 (2004) 1452.
\bibitem{ciu07}P. Ciureanu, L. G. C. Melo, D. Seddaoui, D. Menard, A. Yelon, Physical models of magnetoimpedance,
J. Appl. Phys. 102 (2007) 073908.
\bibitem{cho10}Y. T. Chong, D. Gorlitz, S. Martens, M. Y. E. Yau,S. Allende, J. Bachmann, K. Nielsch,
Multilayered Core/Shell Nanowires Displaying Two Distinct Magnetic Switching Events, Adv. Mater. 22 (2010) 2435.
\bibitem{web12}D. P. Weber, et al., Cantilever Magnetometry of Individual Ni Nanotubes, Nano Lett. 12 (2012) 6139.
\bibitem{str14}R. Streubel, et at., Magnetic Microstructure 
of Rolled-Up Single-Layer Ferromagnetic Nanomembranes, Adv. Mater. 26 (2014) 316. 
\bibitem{gro16}B. Gross, et al., Dynamic cantilever magnetometry of individual CoFeB nanotubes, Phys. Rev. B 93 (2016) 064409. 
\bibitem{wys17}M. Wyss, et al., Imaging magnetic vortex configurations in ferromagnetic nanotubes, Phys. Rev. B 96 (2017) 024423.
\bibitem{sta17}M. Stano, et al., Imaging magnetic flux-closure domains and domain walls in electroless-deposited CoNiB nanotubes, arXiv:1704.06614. 
\bibitem{van01}S. vanDijken, G. Di Santo, B. Poelsema, Influence of the deposition angle on
the magnetic anisotropy in thin Co films on Cu(001), Phys. Rev. B 63 (2001) 104431.
\bibitem{che07}S. Cherifi, et al., Tuning the domain wall orientation in thin magnetic strips using induced anisotropy, Appl. Phys. Lett. 91 (2007) 092502. 
\bibitem{esc07}J. Escrig, P. Landeros, D. Altbir, E. E. Vogel, Effect of anisotropy in magnetic nanotubes, J. Magn. Magn. Mat. 
310 (2007) 2448.
\bibitem{fer17}A. Fernandez-Pacheco, R. Streubel, O. Fruchart, R. Hertel, P. Fischer, R. P. Cowburn,
Three-dimensional nanomagnetism, Nat. Comm. 8 (2017) 15756.
\bibitem{chi09}H. Chiriac, O.-G. Dragos, M. Grigoras, G. Ababei, N. Lupu, Magnetotransport
Phenomena in [NiFe/Cu] Magnetic Multilayered Nanowires, IEEE Trans. Magn. 45 (2009) 4077.
\bibitem{kur04}G. V. Kurlyandskaya, et al., Domain structure and magnetization process
of a giant magnetoimpedance geometry FeNi/Cu/FeNi(Cu)FeNi/Cu/FeNi sensitive element,
J. Phys.: Cond. Matter 16 (2004) 6561.
\bibitem{nak14}K. S. Nakayama, T. Chiba, S. Tsukimoto, Y. Yokoyama, T. Shima, S. Yabukami,
Ferromagnetic resonance in soft-magnetic metallic glass nanowire and microwire, Appl. Phys. Lett. 105 (2014) 202403.
\bibitem{fer10}E. Fernandez, A. Garcia-Arribas, S. O. Volchkov, G. V. Kurlyandskaya, J. M. Barandiaran,
Differences in the Magneto-Impedance of FeNi/Cu/FeNi Multilayers With Open and Closed Magnetic Path,
IEEE Trans. Magn. 46 (2010) 658.
\bibitem{ata13}F. E. Atalay, H. Kaya, S. Atalay, E.Aydogmus, Magnetoimpedance effects in a CoNiFe nanowire array,
J. Alloys and Comp. 561 (2013) 71.
\bibitem{jan16}A. Janutka, Ordering in rolled-up single-walled ferromagnetic nanomembranes,
J. Magn. Magn. Mater. 419 (2016) 282.
\bibitem{bea08}G. S. D. Beach, M. Tsoi, J. L. Erskine, Current-induced domain
wall motion,  J. Magn. Magn. Mater. 320 (2008) 1272.
\bibitem{tse08}Y. Tserkovnyak, A. Brataas, G. E. W. Bauer, Theory of
current-driven magnetization dynamics in inhomogeneous
ferromagnets, J. Magn. Magn. Mater. 320 (2008) 1282. 
\bibitem{gar01}H. Garcia-Miquel, et al., Power Absorption and Ferromagnetic Resonance in
Co-Rich Metallic Glasses, IEEE Trans. Magn. 37 (2001) 561. 
\bibitem{kra03}L. Kraus, GMI modeling and material optimization, Sensors and Actuators A 106 (2003) 187. 
\bibitem{kra99}L. Kraus, Theory of giant magneto-impedance in the planar conductor
with uniaxial magnetic anisotropy, J. Magn. Magn. Mater. 195 (1999) 764.
\bibitem{kit48}C. Kittel, On the theor of ferromagnetic resonance absorption, Phys. Rev. 73 (1948) 155.
\bibitem{dau07}M. Daub, M. Knez, U. Goesele, K. Nielsch, Ferromagnetic nanotubes by atomic layer deposition in anodic alumina membranes, J. Appl. Phys. 101 (2007) 09J111.
\bibitem{mul09}C. Muller, et al.,Tuning magnetic properties by roll-up of Au/Co/Au films into
microtubes, Appl.Phys.Lett. 94 (2009) 102510.
\bibitem{thi07}A. Thiaville, Y. Nakatani, F. Piechon, J. Miltat, T. Ono, Transient 
domain wall displacement under spin-polarized current pulses, Eur. Phys. J. B 60 (2007) 15.
\bibitem{thi06}A. Thiaville, Y. Nakatani, Domain-Wall Dynamics in Nanowires and Nanostrips,
in {\it Spin Dynamics in Confined Magnetic Structures III}, (Springer, Berlin, 2006); pp. 161-206
\bibitem{jan11}A. Janutka, Externally driven transmission and collisions of domain walls
in ferromagnetic wires, Phys. Rev. E 83 (2011) 056607.
\bibitem{sch74}N. L. Schryer, L. R. Walker, The motion of 180 domain walls in uniform dc magnetic fields,
J. Appl. Phys. 45 (1974) 5406.
\bibitem{oommf}http://math.nist.gov/oommf/
\bibitem{ota13}J. A. Otalora, J. A.Lopez-Lopez, P.Landeros, P.Vargas, A.S.Nunez, Breaking of chiral symmetry in vortex domain wall propagation in ferromagnetic nanotubes, J. Magn. Magn. Mater. 341 (20013) 86.
\bibitem{tay05}J. R. Taylor, Classical Mechanics, (University Science Books 2005); chap. 12.
\bibitem{vaz01}M. Vazquez, Giant magneto-impedance in soft magnetic "Wires", 
J. Magn. Magn. Mat. 226-230  (2001) 693.
\bibitem{cla08}D. J. Clarke, O. A. Tretiakov, G.-W. Chern, Ya. B. Bazaliy, O. Tchernyshyov,
Dynamics of a vortex domain wall in a magnetic nanostrip: Application of the
collective-coordinate approach, Phys. Rev. B 78 (2008) 134412. 
\bibitem{mac96}F. L. A. Machado, S. M. Rezende, A theoretical model
for the giant magnetoimpedance in ribbons of amorphous soft-ferromagnetic alloys,
J. Appl. Phys. 79 (1996) 6558.
\bibitem{jan13}A. Janutka, Complexes of Domain Walls in One-Dimensional Ferromagnets Near
and Far from Phase Transition, Acta Phys. Pol. A 124 (2013) 23. 
\bibitem{zhu13}A. Zhukov, M. Ipatov, C. Garcia, M. Churyukanova, S. Kaloshkin, V. Zhukova,
From Manipulation of Giant Magnetoimpedance in Thin Wires to Industrial Applications, 
J. Supercond. Nov. Magn. 26 (2013) 1045. 
\bibitem{elk16}R. El Kammouni, G. V. Kurlyandskaya, M. Vazquez, S. O. Volchkov, Magnetic
properties and magnetoimpedance of short CuBe/CoFeNi electroplated microtubes, 
Sensors and Actuators A 248 (2016) 155. 
\bibitem{zim15a}J. Ziman, M. Kladivova, V. Suhajova, Impedance and domain wall mass determination
in cylindrical wire with circular anisotropy, J. Magn. Magn. Mat. 393 (2015) 363. 
\bibitem{kle17}P. Klein, J. Onufer, J. Ziman, G. A. Badini-Confalonieri, M. Vazquez, R. Varga,
Effect of Annealing on Domain Wall Mass in Amorphous FeCoMoB Microwires, IEEE Trans. Magn. 53 (2017) 4300404.
\bibitem{fan11}H. Fangohr, D. S. Chernyshenko, M. Franchin, T. Fischbacher, G. Meier, Joule heating in nanowires,
Phys. Rev. B 84 (2011) 054437.
\bibitem{thi05}A. Thiaville, Y. Nakatani, J. Miltat, Y. Suzuki,
Micromagnetic understanding of current-driven domain wall motion in patterned nanowires,
Europhys. Lett. 69 (2005) 990. 
\bibitem{bea07}G.S.D. Beach, C Knutson, M. Tsoi, J.L. Erskine, Field- and current-driven domain wall dynamics: An experimental picture,  J. Magn. Magn. Mater. 310 (2007) 2038.
\bibitem{vel94}M. Velazquez, M. Vazquez, D.-X. Chen, A. Hernando, Giant magnetoimpedance
in nonmagnetostrictive amorphous wire, Phys. Rev. B 50 (1994) 16737.
\end{thebibliography}
\end{document}